\documentclass[a4paper,fleqn,usenatbib]{mnras}

\usepackage{newtxtext,newtxmath}

\usepackage[T1]{fontenc}
\usepackage{ae,aecompl}

\usepackage{graphicx}	
\usepackage{amsmath}	
\usepackage{amssymb}	
\usepackage{hyperref}

\def\apj{ApJ}

\def\araa{ARA\&A}
\def\aap{A\&A}
\def\apjl{ApJL}
\def\apjs{ApJS}
\def\mnras{MNRAS}
\def\aj{AJ}
\def\nat{Nature}

\def\icarus{Icarus}

\hypersetup{
  draft=true,
	pdftoolbar=false,
	pdfmenubar=true,
	pdfstartview={FitP},
	breaklinks=false,
	frenchlinks=false,
  colorlinks=true,
  linkcolor=red,
  citecolor=blue,
  filecolor=cyan,
  urlcolor=magenta,
  pdftitle={Exoplanetary Monte Carlo Radiative Transfer with Correlated-k I. Benchmarking Transit and Emission Observables},
	pdfsubject={radiative transfer, planets and satellites: atmospheres, methods: numerical},
	pdfauthor={Graham K. H. Lee, Jake Taylor, Mark Hammond, Jean-Loup Baudino, Ryan Garland, Kenneth Wood and Patrick G. J. Irwin}
  }

\title[Exoplanetary MCRT with Corr-k I.]{Exoplanetary Monte Carlo Radiative Transfer with Correlated-k \\
 I. Benchmarking Transit and Emission Observables}

\author[Lee et al.]{
Graham K. H. Lee$^{1}$\thanks{E-mail: graham.lee@physics.ox.ac.uk},
Jake Taylor$^{1}$,
Simon L. Grimm$^{2}$,
Jean-Loup Baudino$^{1}$, \newauthor
Ryan Garland$^{1}$,
Patrick G. J. Irwin$^{1}$,
and Kenneth Wood$^{3}$
\\
$^{1}$Atmospheric, Oceanic \& Planetary Physics, Department of Physics, University of Oxford, Oxford OX1 3PU, UK \\
$^{2}$Center for Space and Habitability, University of Bern, Gesellschaftsstrasse 6, 3012 Bern, Switzerland \\
$^{3}$SUPA, School of Physics and Astronomy, University of St Andrews, St Andrews KY16 9SS, UK.
}

\date{Accepted XXX. Received YYY; in original form ZZZ}

\pubyear{2019}

\begin{document}
\label{firstpage}
\pagerange{\pageref{firstpage}--\pageref{lastpage}}
\maketitle

\begin{abstract}
Current observational data of exoplanets are providing increasing detail of their 3D atmospheric structures.
As characterisation efforts expand in scope, the need to develop consistent 3D radiative-transfer methods becomes more pertinent as the complex atmospheric properties of exoplanets are required to be modelled together consistently. \\
We aim to compare the transmission and emission spectra results of a 3D Monte Carlo Radiative Transfer (MCRT) model to contemporary radiative-transfer suites. \\
We perform several benchmarking tests of a MCRT code, Cloudy Monte Carlo Radiative Transfer (CMCRT), to transmission and emission spectra model output.
We add flexibility to the model through the use of k-distribution tables as input opacities.
We present a hybrid MCRT and ray tracing methodology for the calculation of transmission spectra with a multiple scattering component. \\
CMCRT compares well to the transmission spectra benchmarks at the 10s of ppm level.
Emission spectra benchmarks are consistent to within 10\% of the 1D models.
We suggest that differences in the benchmark results are likely caused by geometric effects between plane-parallel and spherical models.
In a practical application, we post-process a cloudy 3D HD 189733b GCM model and compare to available observational data. \\
Our results suggest the core methodology and algorithms of CMCRT produce consistent results to contemporary radiative transfer suites.
3D MCRT methods are highly suitable for detailed post-processing of cloudy and non-cloudy 1D and 3D exoplanet atmosphere simulations in instances where atmospheric inhomogeneities, significant limb effects/geometry or multiple scattering components are important considerations.\\
\end{abstract}

\begin{keywords}
radiative transfer -- planets and satellites: atmospheres -- planets and satellites: individual: HD 189733b -- methods: numerical
\end{keywords}



\section{Introduction}

Observing and characterising the three-dimensional atmospheric structure of exoplanets is a continuing endeavour for the exoplanetary community.
A keystone tool for investigating exoplanet atmospheres in forward and retrieval efforts is radiative-transfer modelling, examining how radiation interacts with the atmosphere and gives rise to the observable properties of each individual exoplanet.

Exoplanet atmospheres are continuing to be observed in ever greater detail.
Transmission spectroscopy from the ground \citep[e.g.][]{Chen2017, Gibson2017, Kirk2017} and space \citep[e.g.][]{Sing2016} has shown many hot and warm Jupiters and Neptunes to contain a variety of molecular and atomic species, for example, Na \citep{Charbonneau2002}, K \citep{Sing2011a}, H$_{2}$O \citep{Swain2009}, CH$_{4}$ \citep{Swain2008} and CO \citep{Snellen2010}, revealing the atmospheric composition at the transmission limbs of the exoplanet atmosphere \citep[e.g.][]{Barstow2017,Tsiaras2018,Fisher2018}.
Many exhibit evidence of cloud coverage at the terminator limbs due to an observed optical wavelength Rayleigh-like slope \citep[e.g.][]{Pont2013}, muted IR water vapour features \citep[e.g.][]{Deming2013} or consistent with a flat, featureless \citep[e.g.][]{Kreidberg2014, Wakeford2017a} grey spectra.
Emission \citep[e.g.][]{Knutson2012} and reflection \citep[e.g.][]{Evans2013} observations have provided information on the dayside temperature structure, and suggest the presence in some exoplanetary atmospheres of a significant optical wavelength scattering component larger than the gas phase Rayleigh scattering component.

3D modelling efforts using Global Circulation Models (GCMs) \citep[e.g.][]{Showman2009, Rauscher2012, Dobbs-Dixon2013, Mayne2014, Mendonca2016} of hot Jupiters and Neptunes suggest atmospheric inhomogeneities in temperature, velocity fields, and vertical mixing rates \citep[e.g.][]{Parmentier2013, Charnay2015a, Kataria2016}.
Gas phase chemistry modelling in 1D \citep[e.g.][]{Drummond2016, Tsai2017, Blumenthal2018} and 2D/3D \citep[e.g.][]{Agundez2012,   Drummond2018, Drummond2018b, Mendonca2018, Steinrueck2018} suggest that considering kinetic non-equilibrium chemistry on the 3D chemical composition of exoplanet atmospheres is important to the resulting temperature structures and observational properties of each individual exoplanet.

Cloud modelling efforts in 1D \citep[e.g.][]{Morley2013,Helling2016, Lavvas2017, Ohno2017, Gao2018, Gao2018b, Powell2018} and 3D \citep[e.g.][]{Charnay2015b, Lee2016, Parmentier2016, Lewis2017, Roman2017, Lines2018, Roman2019} of a variety of sophistications also suggest inhomogeneous cloud coverage in latitude, longitude and depth across the globe of their atmospheres.

With the advent of space missions with dedicated exoplanet atmospheric characterisation payloads, such as the James Webb Space Telescope (JWST) \citep[e.g.][]{Bean2018}, the Wide Field Infra-Red Space Telescope (WFIRST) \citep[e.g.][]{Robinson2016}, and the Atmospheric Remote-sensing Infrared Exoplanet Large-survey (ARIEL) \citep{Tinetti2016}, the need for accurate radiative-transfer modelling for these 3D inhomogeneous objects will become increasingly pertinent for the physical interpretation of exoplanetary observables.
Recently, \citet{Feng2016} and \citet{Blecic2017} have shown the impacts of considering 3D temperature structures when interpreting the retrieval results of emission spectra data.

Through these observational and modelling efforts it is clear that exoplanet atmospheres are inherently 3D and inhomogeneous in nature, with unique local dynamical, temperature, chemical and cloud properties.
Due to the stochastic and microphysical nature of the Monte Carlo Radiative Transfer (MCRT) technique, it is well suited to modelling this complex 3D environment.
In the astrophysical community, MCRT modelling techniques have been proven to be highly successful and constitute the majority of in use 3D radiative-transfer models \citep{Steinacker2013}.
For exoplanets, MCRT has been applied for calculations of transmission spectra \citep{deKok2012, Robinson2017}, geometric albedo and albedo spectra \citep{Hood2008, Munoz2015, Munoz2017} plus emission spectra \citep{Lee2017, Stolker2017}.
In addition, modelling photon processes such as polarization \citep{Stolker2017} and atmospheric refraction \citep{Robinson2017} can be readily included.

In this study, we systematically compare and benchmark the results of the 3D MCRT code Cloudy Monte Carlo Radiative Transfer (CMCRT) presented in \citet{Lee2017} to several contemporary radiative transfer codes, in particular the NEMESIS radiative transfer suite \citep{Irwin2008}, the 3D MCRT model ARTES \citep{Stolker2017} and the \citet{Baudino2017} benchmark protocols.
In Section \ref{sec:MCRT_methods}, we outline updates to the \citet{Lee2017} model and describe aspects of using k-distribution opacities with MCRT.
Sections \ref{sec:trans_spec} and \ref{sec:em_spec} detail our approaches for calculating transmission spectra and emission spectra using MCRT respectively.
In Sect. \ref{sec:benchmarking} we present a direct comparison of the MCRT results to the NEMESIS suite.
In Sect. \ref{sec:Stolker} we present the benchmarking to the emission spectra results of \citet{Stolker2017}.
Section \ref{sec:Baudino} presents the benchmarking to the \citet{Baudino2017} protocols.
Section \ref{sec:GCM} applies our new methodologies to post-process output of the cloudy HD 189733b GCM simulation of \citet{Lee2016}.
Section \ref{sec:discussion} contains the discussion and Sect. \ref{sec:conclusion} contains the conclusions.

\section{MCRT Methods Using K-distributions}
\label{sec:MCRT_methods}

\begin{figure} 
   \centering
   \includegraphics[width=0.49\textwidth]{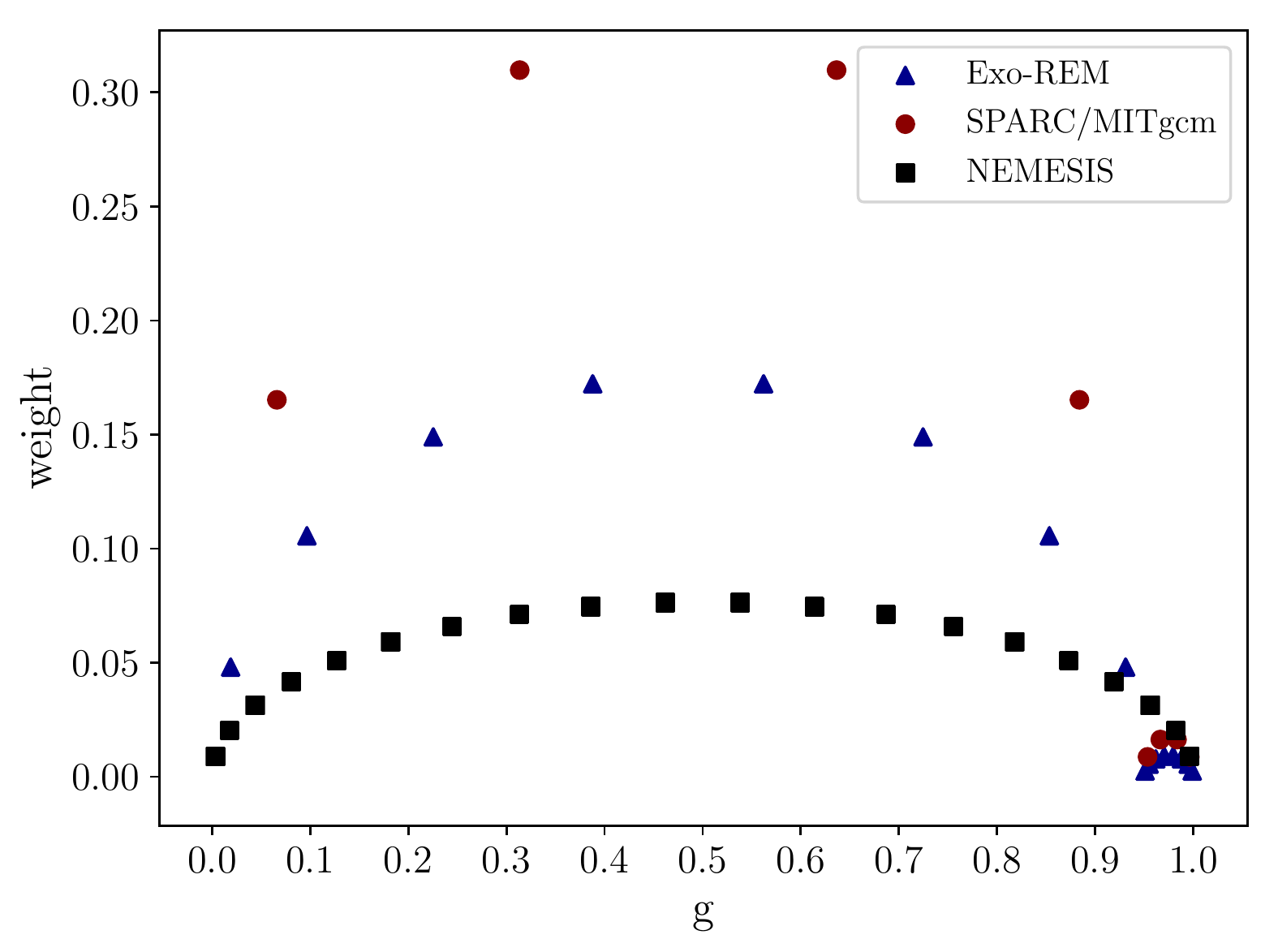}
   \caption{g ordinance values and associated weights for three different RT suites.
   Black squares: (20 points) NEMESIS \citep{Irwin2008}, blue triangles: (8+8 points) Exo-REM \citep{Baudino2015,Baudino2017}, red circles: (4+4 points) SPARC/MITgcm \citep{Showman2009}.}
   \label{fig:g_ord_weight}
\end{figure}

We update and expand the MCRT model described in \citet{Lee2017}, based on the original work by \citet{Hood2008}.
In \citet{Lee2017} pre-mixed \citet{Sharp2007} mean absorption coefficient tables were used.
The MCRT methods are generalised to include the use of individual and mixes of gas species with k-distribution tables for application in exoplanet atmosphere radiative transfer problems.

In the k-distribution method, a high fidelity wavelength or wavenumber absorption coefficient table is re-ordered by its cumulative distribution counterpart for a pre-defined bin width.
A number of points are then sampled from the cumulative distribution function (k-coefficients), given by the values of a Gaussian ordinance (g-ordinance).
Each point is then assigned a weight given by the g-ordinance method used.
However, wavelength information is scrambled in the process, leading to the assumption, for an inhomogeneous media, that the k-coefficients at each Gaussian ordinance are correlated.
This assumption leads to an error when modelling an inhomogeneous atmosphere, as the opacity distribution shifts temperature and pressure dependently.
This error has been shown to retain accuracy to within 10\% when compared to line-by-line tests \citep[e.g.][]{Amundsen2014}.

The advantage of the k-distribution method is that it significantly reduces the computational burden by orders of magnitude compared to line-by-line calculations \citep{Heng2017b}, and is a standard methodology in the planetary science community.
More in-depth descriptions of k-distributions and correlated-k in exoplanet contexts can be found in \citet{Amundsen2014, Grimm2015, Heng2017b} and \citet{Amundsen2017}.
When required, we apply the random overlap method \citep[e.g.][]{Lacis1991,Amundsen2017} to combine individual gas k-tables weighted by their relative abundances.

Due to the statistical nature of both MCRT and k-coefficients, they have very complimentary properties.
In MCRT, each packet is evolved in the simulation by sampling a normalised cumulative distribution function (CDF), $\psi$, of a probability function, $p(x)$, from the general equation \citep[e.g.][]{Stolker2017}
\begin{equation}
\psi(x_{0}) = \frac{\int_{a}^{x_{0}}p(x) dx}{\int_{a}^{b}p(x) dx},
\end{equation}
where $x_{0}$ is a randomly sampled variable, with $a$ and $b$ the lower and upper limit of the distribution.

MCRT can make use of k-tables by sampling the weighted cumulative distribution properties of the k-distribution method.
After a packet is spawned in the simulation for a given wavelength bin, a g-ordinance can be randomly selected for that packet by sampling the cumulative distribution function of the Gaussian quadrature weights, w$_{g}$, where the probability of sampling a specific g-ordinate, $g$, is
\begin{equation}
\label{eq:g_samp}
g_{samp} = \frac{w_{g}}{\sum_{g} w_{g}}.
\end{equation}
In the MCRT simulation, a uniformly sampled random number, $\zeta$ $\in$ [0,1], is drawn for each packet which corresponds to a single g-ordinance value.
This g-ordinance value is then retained throughout the packet's lifetime, and used for all further calculations involving the packet.
The correlated-k approximation \citep[e.g.][]{Heng2017b} is therefore inherent since we are assuming that the g-ordinance sampled for the packet is correlated across the entire 3D atmosphere.

To visualise the sampling of the g-ordinance, Fig. \ref{fig:g_ord_weight} shows examples of the g values and weights from numerous correlated-k approaches in the literature:
\begin{itemize}
\item The uniform 20 ordinance points typically used by the NEMESIS \citep{Irwin2008} radiative-transfer suite .
\item The 4+4 used in the SPARC/MITgcm \citep{Showman2009} GCM model.
\item The 8+8 used in the Exo-REM \citep{Baudino2015,Baudino2017} radiative-convective equilibrium model.
\end{itemize}
In typical use, the additional higher order ordinance points are used in GCM and radiative-convective modelling due to their larger bin sizes \citep[e.g.][]{Showman2009, Amundsen2014}.
This is performed in an attempt to capture the opacity contribution of numerous line centers in each bin, while keeping the efficiency of the radiative-transfer scheme reasonable.
For retrieval and post-processing efforts with smaller bin sizes (hence less numbers of line centers), a uniform spacing with a larger number of g-ordinances is sufficient to capture the opacity distribution well.
In traditional correlated-k radiative transfer codes, the Gaussian quadrature procedure is applied to calculate integrated mean quantities for a given wavelength bin.

In the context of MCRT, Eq. \ref{eq:g_samp} shows that larger weighted g-ordinance values are more likely to be sampled by the scheme.
Since the weights are normally distributed, central g-ordinance points are more likely to be sampled compared to those at the distribution wings.
The MCRT scheme therefore acts as a numerical integrator by sampling the weighted distribution and summing the results of each sample at the simulation end.
As with any Monte Carlo integration calculation, less variance in the final answer is obtained by increasing the number of samples.
In principle, any number or sampling scheme of g-ordinances with weights can be used in the MCRT scheme.
Using less g-ordinances will not necessarily decrease the computational run-time in this model, since the total number of sampled packets primarily controls the run-time.
Individual wavelength calculations are retained by using a single g-ordinate with unity weight.

\section{Transmission Spectra}
\label{sec:trans_spec}

\begin{figure} 
   \centering
   \includegraphics[width=0.49\textwidth]{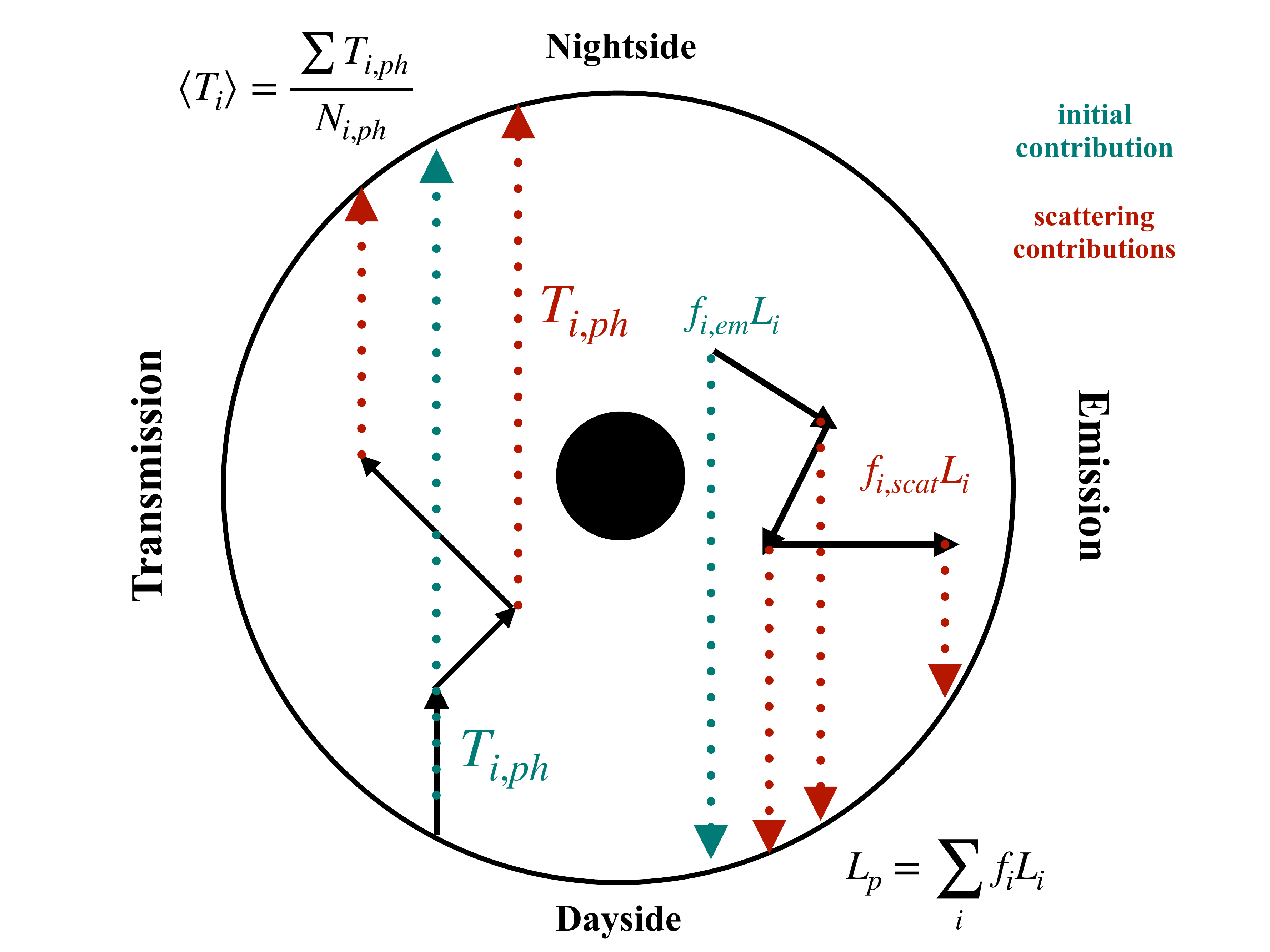}
   \caption{Schematic diagram visualising the transmission (LHS) and emission (RHS) spectra modes of CMCRT.
   CMCRT is a hybrid scheme, combining a ray tracing method (dotted, coloured lines) with the evolution of individual packets (black arrows) to model observable quantities from the 3D spherical grid.}
   \label{fig:CMCRT_scheme}
\end{figure}

Stellar light travelling through a planetary atmosphere principally interacts through the extinction of photons in the line of sight direction.
This extinction reduces the average transmission of stellar light through the atmosphere and imprints atomic and molecular features in the transmission spectra.
In addition, photons may also undergo multiple scattering interactions in the atmosphere, and subsequently escape towards the line of sight.
Multiple scattered photons exiting the atmosphere towards the line of sight will increase the average transmission compared to extinction alone, resulting in the planet appearing smaller than it would without a multiple scattering component.
This brightening effect from multiple scattered photons was first studied in the context of exoplanet transmission spectra by \citet{Hubbard2001}.

The transmission spectrum formula in discretised form is given by \citep[e.g.][]{Dobbs-Dixon2013, Robinson2017}
\begin{equation}
\label{eq:transit_form}
\left(\frac{R_p}{R_\star}\right)^2 = \frac{1}{R_\star^2}\left(R_{p,0}^2 + 2\sum_{i=1}^{N_{i}} [1 - \langle T_{i}\rangle]b_{i}\Delta b_{i}\right),
\end{equation}
where $R_{p}$ [cm] is the apparent radius of the planet, $R_{\star}$ [cm] the radius of the host star, $R_{p, 0}$ [cm] defined as the radius below which the planet can be considered an opaque solid body, $\langle T_{i}\rangle$ the mean transmission at impact parameter index $i$, $b_{i}$ [cm] the height of the impact parameter chord and $\Delta b_{i}$ [cm] the radial width of transit chord $i$.

Central to the CMCRT transmission spectra model is calculating the transmission through the atmosphere at each transit chord including the effects of multiple scattering.
The sequence of events for a packet in the transmission spectrum calculation is as follows:
\begin{enumerate}
\item A packet is spawned at a random impact parameter at the transmission annulus of the 3D spherical grid.
\item The packet is assigned a g-ordinate value, by randomly sampling the cumulative k-distribution weights (Eq. \ref{eq:g_samp}).
\item An initial transmission calculation towards the observational direction is performed.
\item The packet is evolved through the simulation grid until termination or escape.
\item The packet's contribution to the transmission from scattering events is recorded throughout its lifetime through the use of the next event estimation method.
\end{enumerate}
Figure \ref{fig:CMCRT_scheme} (LHS) shows a diagrammatic representation of the path of a photon packet and the key steps considered in the simulation.

In our hybrid scheme, we combine the MCRT with a ray tracing method, the next event estimation technique \citep{Yusef-Zadeh1984,Wood1999}, to calculate the contribution of each packet to the transmission through each impact parameter.
The initial transmission contribution of a single packet, $T_{i, ph}$, initially at a random location across the impact parameter width of vertical index $i$ is
\begin{equation}
T_{i, ph} =  \exp(-\tau_{i, ph}),
\end{equation}
where $\tau_{i, ph}$ is the total optical depth towards the simulation boundary in the observational direction.
At each scattering location for the packet during its lifetime, the scattering event also contributes an additional transmission to the packet's current impact parameter index $i$ given by
\begin{equation}
T_{i, ph} =  \omega W_{ph} \Phi(\alpha) \exp(-\tau_{i, ph}),
\end{equation}
where $\omega$ is the local single scattering albedo, $W_{ph}$ is the current weight of the packet and $\Phi(\alpha)$ the normalised scattering phase function probability towards observation direction $\alpha$.
It is important that this scattering contribution is calculated at the impact parameter of the current 3D location of the packet, which may be different from its originally initialised impact parameter after many scattering events.
The mean transmission through impact parameter index $i$, $\langle T_{i}\rangle$, is then the total transmission from all contributing packets (initial plus scattering contributions), normalised by the number of packets that was originally randomly initialised across the impact parameter index, $N_{i, ph}$,
\begin{equation}
\langle T_{i}\rangle = \frac{\sum T_{i, ph}}{N_{i, ph}}.
\end{equation}
This mean transmission at each impact parameter index is then used in Eq. \eqref{eq:transit_form} to calculate the radius ratio.

In summary, all simulated packets contribute a transmission through an impact parameter vertical index $i$, which is tracked for all packets throughout the simulation runtime.
Since we are using the next event estimation method, the end points of the packets do not determine the transmission spectrum, and every packet contributes to the final solution.
This increases the overall efficiency, and helps lower the variance of the model (e.g. see discussion in \citet{Lee2017}).
We also apply the survival biasing with Russian Roulette scheme \citep[e.g.][]{Dupree2002, Lee2017} in order to more accurately capture the effects of multiple scattering on the transmission spectra.

A feature of the model is that should the atmosphere have zero scattering opacity, or packet scattering be turned off, the scheme reduces to a 3D extinction limit ray tracing model.
The impact of individual scattering components on the end spectra can be examined by running the simulation with and without packet scattering.

\section{Emission Spectra}
\label{sec:em_spec}

For emission spectra calculations we use the same scheme as in \citet{Lee2017} but with some additional properties when using k-distribution tables.
The monochromatic luminosity, $L_{i, \lambda}$ [erg s$^{-1}$ cm$^{-1}$], of cell $i$ is given by \citep[e.g.][]{Pinte2006, Stolker2017}
\begin{equation}
\label{eq:L_i}
L_{i, \lambda} = 4\pi\rho_{i}V_{i}\kappa_{i,\lambda}^{abs}B_{\lambda}(T_{i}),
\end{equation}
where $\rho_{i}$ [g cm$^{-3}$] is the cell gas density, $V_{i}$ [cm$^{3}$] the cell volume, $\kappa_{i,\lambda}^{abs}$ [cm$^{2}$ g$^{-1}$] the mass absorption opacity (possibly including cloud particle absorption opacity) and $B_{\lambda}(T_{i})$ [erg s$^{-1}$ sr$^{-1}$ cm$^{-3}$] the Planck function at temperature $T_{i}$ [K].

Using k-distribution coefficients, the luminosity contribution from each g-ordinate in the wavelength bin is given by
\begin{equation}
\label{eq:L_g}
L_{i, g} = 4\pi\rho_{i}V_{i}\kappa_{i, g}^{abs}B_{\lambda}(T_{i}),
\end{equation}
where $\lambda$ [cm] is taken as the bin center wavelength, with the total pseudo-spectral luminosity in a band $\Delta\lambda$ given by
\begin{equation}
\label{eq:L_g_samp}
L_{i, \Delta\lambda} = \sum_{g}w_{g}L_{i, g}.
\end{equation}
The probability of sampling a particular g-ordinate for the photon packet in each cell is now given by
\begin{equation}
\label{eq:gl_samp}
g_{samp} = \frac{w_{g}L_{i, g}}{\sum_{g}w_{g}L_{i ,g}},
\end{equation}
which has the important property that the emission spectra in each wavelength bin is more strongly determined by the higher order g-ordinances, corresponding to greater sampling of (near) line-centers in the wavelength bin.
The total luminosity of the planet at a given wavelength bin, $L_{p,\Delta\lambda}$ [erg s$^{-1}$], is the sum of the luminosity of each cell, multiplied by the fraction of energy that escaped from that cell, $f_{i}$, toward the observational direction through the next event estimation method \citep{Lee2017},
\begin{equation}
L_{p,\Delta\lambda} = \sum_{i} f_{i}L_{i, \Delta\lambda}.
\end{equation}

The sequence of events for an emission spectra packet is as follows:
\begin{enumerate}
\item A packet is spawned at a random starting position within a cell volume.
\item A g-ordinance is sampled for the packet given by Eq. \eqref{eq:gl_samp}.
\item An initial transmission calculation towards the observational direction is performed.
\item The packet is evolved through the simulation until termination or escape.
\item The fraction of energy escaping towards the observational direction through scattering events is tracked throughout the packet's lifetime through the next event estimation method.
\end{enumerate}
Figure \ref{fig:CMCRT_scheme} (RHS) shows a diagrammatic representation of the emission scheme in CMCRT.

An assumption in this approach is that Eq. \eqref{eq:L_g} implies that the value of $B_{\lambda}(T_{i})$ is constant across the wavelength bin range, and equal to the bin center wavelength.
This approximation is reasonable when considering the small bin sizes ($<$ 0.05 $\mu$m) of the current post-processing efforts.
However, this approximation may be invalid for large bin sizes where the Planck function would vary significantly across the bin edges.

Lastly, we note that our emission spectra model accounts for the wavelength dependent photospheric radius effect discussed in \citet{Fortney2019}.
The wavelength dependence of the photospheric radius is self-consistency accounted for in the 3D grid, since the fraction of a cell's luminosity, directly weighted by the available emitting material in the cell (Eq. \ref{eq:L_i}), escaping toward the observational direction is calculated.
Additionally, any variation in the emitting area as a function of planetary orbital phase is readily accounted for.

\section{Benchmarking Tests}
\label{sec:benchmarking}

\begin{figure} 
   \centering
   \includegraphics[width=0.49\textwidth]{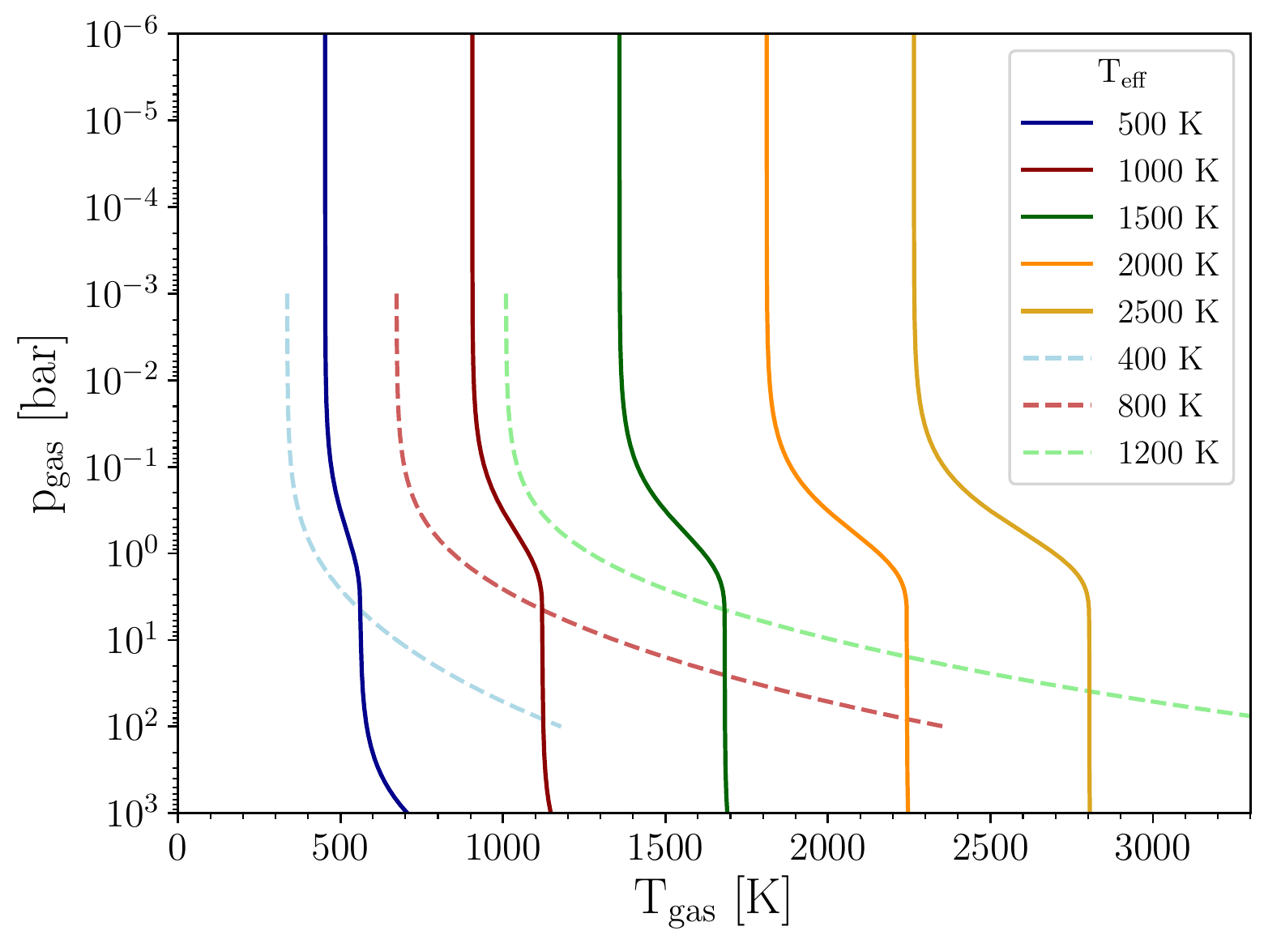}
   \caption{Temperature-pressure profiles used in this study.
   Dashed lines denote the benchmark profiles from \citet{Stolker2017} (Sect. \ref{sec:Stolker}), solid lines denote the benchmark profiles from \citet{Baudino2017} (Sect. \ref{sec:Baudino}).}
   \label{fig:TP}
\end{figure}

We perform all simulations on a 3D spherical grid with longitude and latitude sizes ($\theta$, $\phi$) = (141, 61) corresponding to a typical exoplanet GCM simulation grid resolution \citep[e.g.][]{Dobbs-Dixon2013,Mayne2014,Kataria2016}.
The radial grid size is variable depending on the benchmark protocols.
All temperature-pressure profiles used in this study are presented in Fig. \ref{fig:TP}.

The MCRT model uses a radial height based spherical coordinate grid for all calculations.
We apply the height output from the NEMESIS code directly as the grid for the NEMESIS benchmark tests.
For all other tests, a second order hydrostatic calculator (benchmarked to the NEMESIS output) is used to calculate the altitude given a temperature-pressure structure.

For input gas phase opacities we use k-distribution tables produced by the NEMESIS \citep{Irwin2008} and HELIOS-K \citep{Grimm2015} opacity calculators, dependent on the benchmark.
All benchmarks presented here were repeated using both sets of k-tables, with only minor differences found between the output, mainly stemming from the different wavelength/wavenumber resolutions used for each opacity set.
The input line-lists are presented in Table \ref{tab:opac_sources}.

We use a custom made opacity tool which reads in individual k-tables, performs the T-p grid interpolation and random overlap k-table mixing, and is responsible for generating the opacity structure of the model atmosphere to be read in by CMCRT.
This code also calculates any required gas phase Rayleigh scattering opacities, continuum opacities and cloud optical properties.

When applying the opacities produced by NEMESIS, we use an evenly spaced wavelength resolution of $\Delta$$\lambda$ = 0.005 $\mu$m for 5929 wavelength bins between 0.305 $\mu$m and 29.945 $\mu$m.
Each bin is sampled using 20 uniform g-ordinance points (Fig. \ref{fig:g_ord_weight}).
Several tests were conducted with 5$\times$ ($\Delta$$\lambda$ = 0.001 $\mu$m) and a tenth ($\Delta$$\lambda$ = 0.05 $\mu$m) this wavelength resolution for both the emission and transmission spectrum modes, all producing consistent results.
This suggests a low-sensitivity of the results for these wavelength bin widths.
For the HELIOS-K opacities, we use an evenly spaced grid of $\Delta$$\nu$ = 10 cm$^{-1}$ between 0-30000 cm$^{-1}$ wavenumber.
We use the Chebyshev polynomial opacity distribution fits from HELIOS-K \citep{Grimm2015}, sampling 20 uniform g-ordinance points for each bin, the same as the NEMESIS tables.
We currently do not include Na and K gas phase opacities in the HELIOS-K benchmarks, as they are not yet included in the HELIOS-K database.

\begin{table}
\caption{Gas phase absorbers and Rayleigh scattering species used as opacity sources in this study.
The current HELIOS-K tables do not include Na and K opacities.}
\begin{center}
\begin{tabular}{l l}  \hline \hline
Species & Reference: NEMESIS - HELIOS-K \\ \hline
H$_{2}$O & \citet{Barber2006} - \citet{Polyansky2018} \\
CH$_{4}$ & \citet{Yurchenko2014} \\
CO & \citet{Rothman2010} - \citet{Li2015} \\
CO$_{2}$ & \citet{Tashkun2011} - \citet{Rothman2010} \\
NH$_{3}$ & \citet{Yurchenko2010} \\
PH$_{3}$ & \citet{Sousa-Silva2015} \\
Na & \citet{Heiter2008} -  N/A\\
K & \citet{Heiter2008} - N/A \\
H$_{2}$-H$_{2}$ CIA &  \citet{Richard2012} \\
H$_{2}$-He CIA &  \citet{Borysow2001, Borysow2002} \\
\hline
Scattering &  \\ \hline
H$_{2}$ & \citet{Irwin2009} \\
He & \citet{Irwin2009} \\
\end{tabular}
\end{center}
\label{tab:opac_sources}
\end{table}%

\subsection{NEMESIS Benchmarking}
\label{sec:NEMESIS}

\begin{figure*} 
   \centering
   \includegraphics[width=0.48\textwidth]{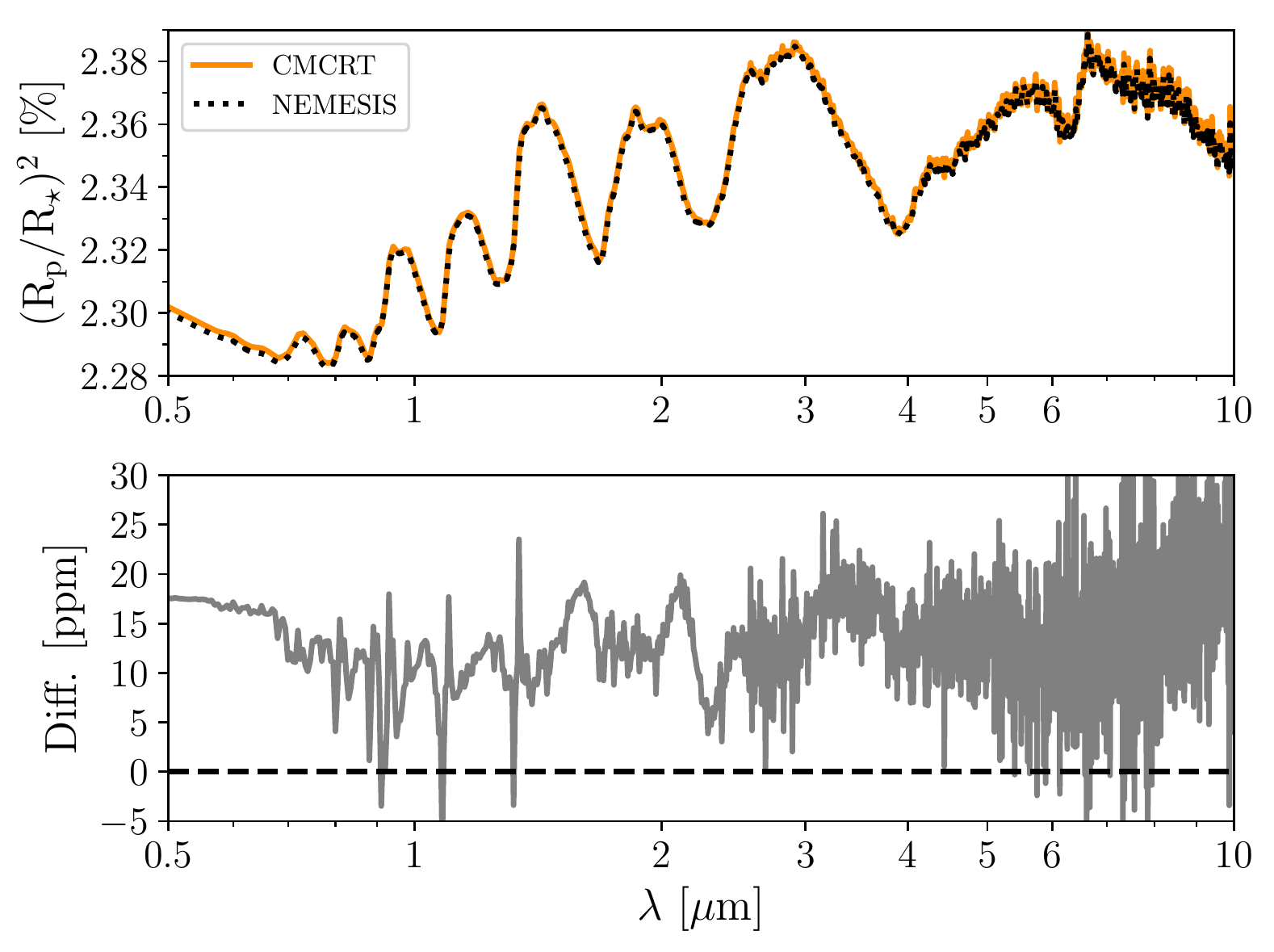}
   \includegraphics[width=0.48\textwidth]{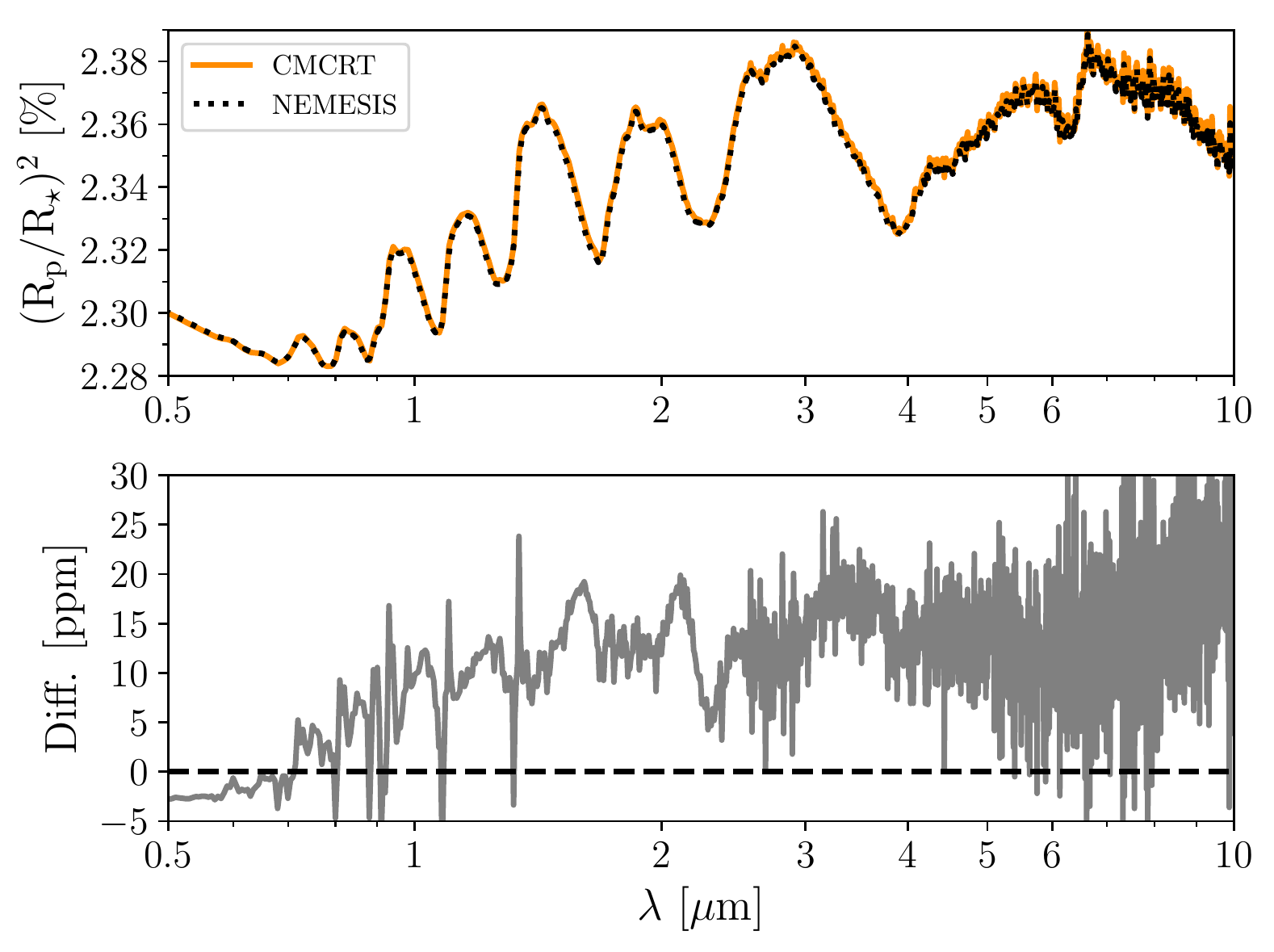}
   \caption{Comparison between the NEMESIS and CMCRT transmission spectrum results for the HD 189733b-like benchmark.
   Upper panels show the transmission spectrum of each model (CMCRT; dark-orange solid line, NEMESIS; black dotted line).
   Lower panels show the difference (CMCRT minus NEMESIS) between the models in ppm.
   Left: CMCRT without multiple scattering (equivalent extinction).
   Right: CMCRT allowing the multiple scattering component.}
   \label{fig:Trans_1500K_NEMESIS}
\end{figure*}

\begin{figure*} 
   \centering
   \includegraphics[width=0.48\textwidth]{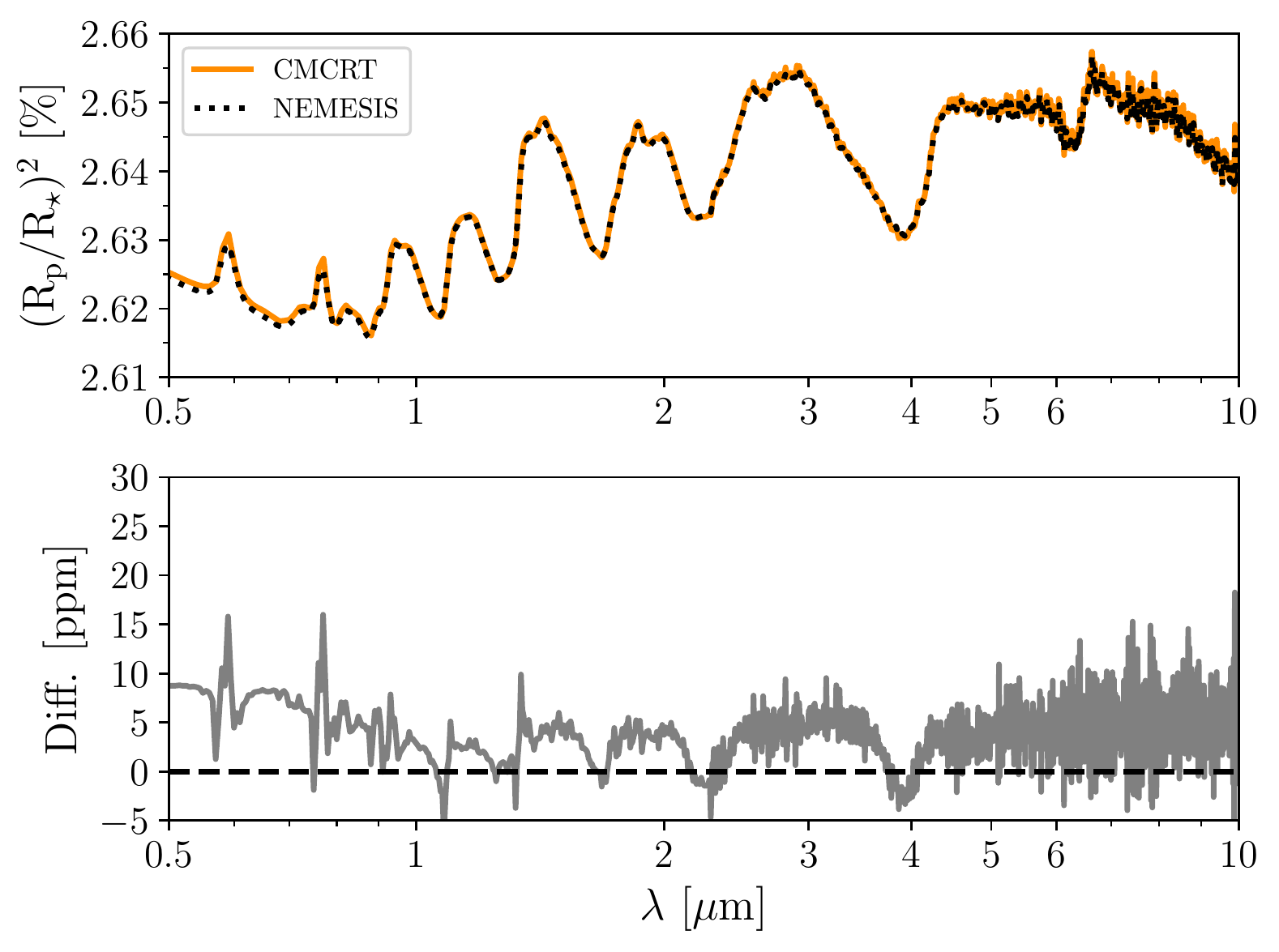}
   \includegraphics[width=0.48\textwidth]{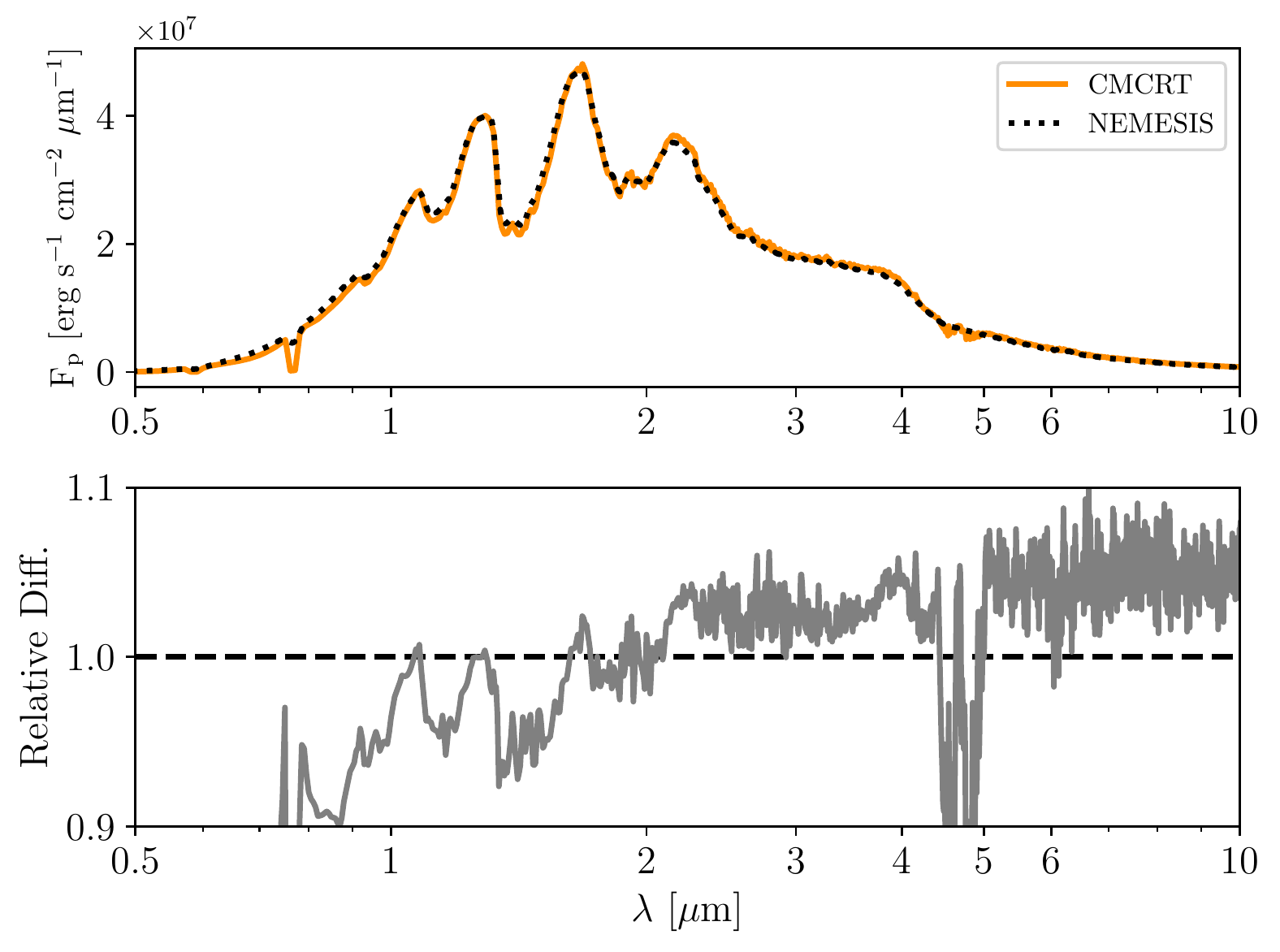}
   \caption{Comparison between the NEMESIS and CMCRT transmission and emission spectrum calculation for the T$_{\rm eff}$ = 1500 K benchmark case (Sect. \ref{sec:Baudino}).
   Upper: Transmission (Left) and emission (Right) spectrum of each model (CMCRT; dark-orange solid line, NEMESIS; black dotted line).
   Lower: Difference (Left: CMCRT minus NEMESIS) and relative difference (Right: CMCRT $\div$ NEMESIS) between the models.}
   \label{fig:Em_1500K_NEMESIS}
\end{figure*}

In our first benchmark test, we perform a direct comparison between CMCRT and the NEMESIS radiative-transfer suite of \citet{Irwin2008}.
We use the same input values (temperatures, pressures, heights, opacities etc) as to allow a direct comparison between the core algorithms.
For input values, the commonly used HD 189733b-like benchmark from \citet{Robinson2017}, with bulk planetary parameters $R_{p, 0}$ (10 bar) = 1.16 $R_{j}$ (1 R$_{j}$ = 6.9911 $\cdot$ 10$^{9}$ cm), M$_{p}$ = 1.14 M$_{j}$ (1 M$_{j}$ = 1.89813 $\cdot$ 10$^{30}$ g) is used.
The atmospheric properties are 156 layers between 10$^{-9}$-10 bar in log-space gas pressure, 1500 K isothermal gas temperatures and constant mean molecular weight $\bar{\mu}$ = 2.316 g mol$^{-1}$.
A constant volume mixing ratio of 0.85, 0.15 and 4 $\cdot$ 10$^{-4}$ is applied for H$_{2}$, He and H$_{2}$O respectively.
The input opacity sources are H$_{2}$-H$_{2}$ and H$_{2}$-He collisional induced absorption, H$_{2}$ and He Rayleigh scattering and H$_{2}$O absorption.

Figure \ref{fig:Trans_1500K_NEMESIS} presents the results of the transmission spectra benchmark.
CMCRT compares well to the NEMESIS output, with the differences between the models remaining below 30 ppm for the equivalent extinction case.
A systematic positive offset of $\approx$10-30 ppm is seen across the infrared wavelengths.
When multiple scattering is allowed in CMCRT, the increased transmission drops the optical Rayleigh slope by $\approx$ 20 ppm.
Since NEMESIS includes multiple-scattering in its formalisms \citep{Barstow2014}, the better agreement with CMCRT in the multiple-scattering case is a promising indication that the scattering contributions are being accurately captured in CMCRT.

As a test of a more realistic atmospheric environment, we produced transmission and emission spectra of the T$_{\rm eff}$ = 1500 K benchmark case from Sect. \ref{sec:Baudino} (Fig. \ref{fig:Em_1500K_NEMESIS}).
The transmission spectra agree to within $\approx$10 ppm, with another small $\approx$5 ppm systematic positive offset in the infrared regions.
The emission spectra also agrees well, with a relative difference to within 10\% across the wavelength range.
Slightly deeper Na, K, H$_{2}$O and CO absorption features are produced by CMCRT compared to the NEMESIS spectra, suggesting a different spectral line formation region present in the 3D CMCRT grid.

The high frequency variations between the models are attributed to the Monte Carlo noise error of sampling the g-ordinance weights, evidenced by the direct imprint of absorption opacity features in the lower panels of Fig. \ref{fig:Trans_1500K_NEMESIS} and Fig. \ref{fig:Em_1500K_NEMESIS}.
However, it is clear a small systematic offset is seen between the models for both the transmission and emission spectra results.
In Sect. \ref{sec:discussion} we discuss possible geometric differences as an explanation for the systematic offsets.

\subsection{Stolker et al. benchmark}
\label{sec:Stolker}

\begin{figure} 
   \centering
   \includegraphics[width=0.48\textwidth]{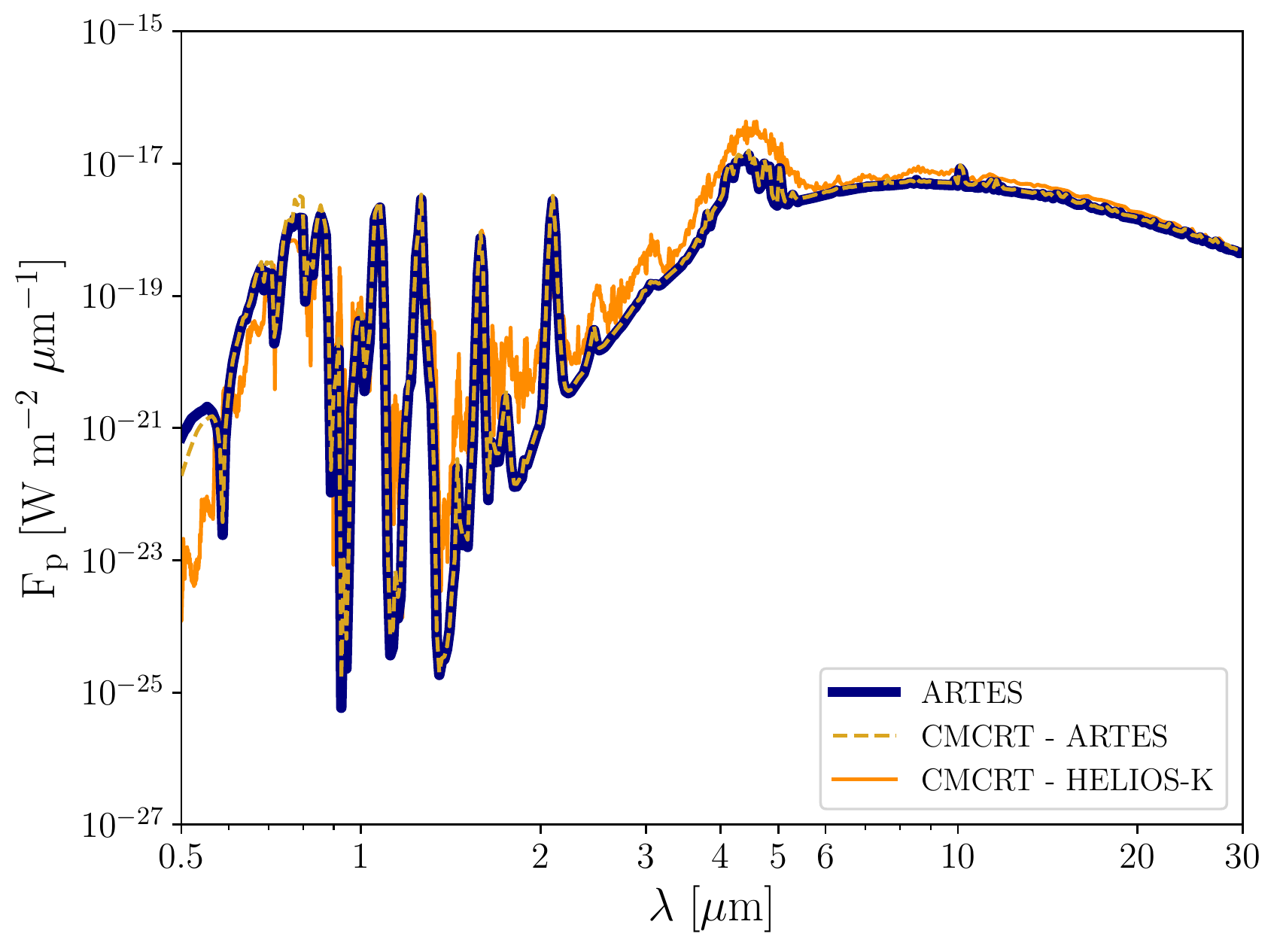}
   \includegraphics[width=0.48\textwidth]{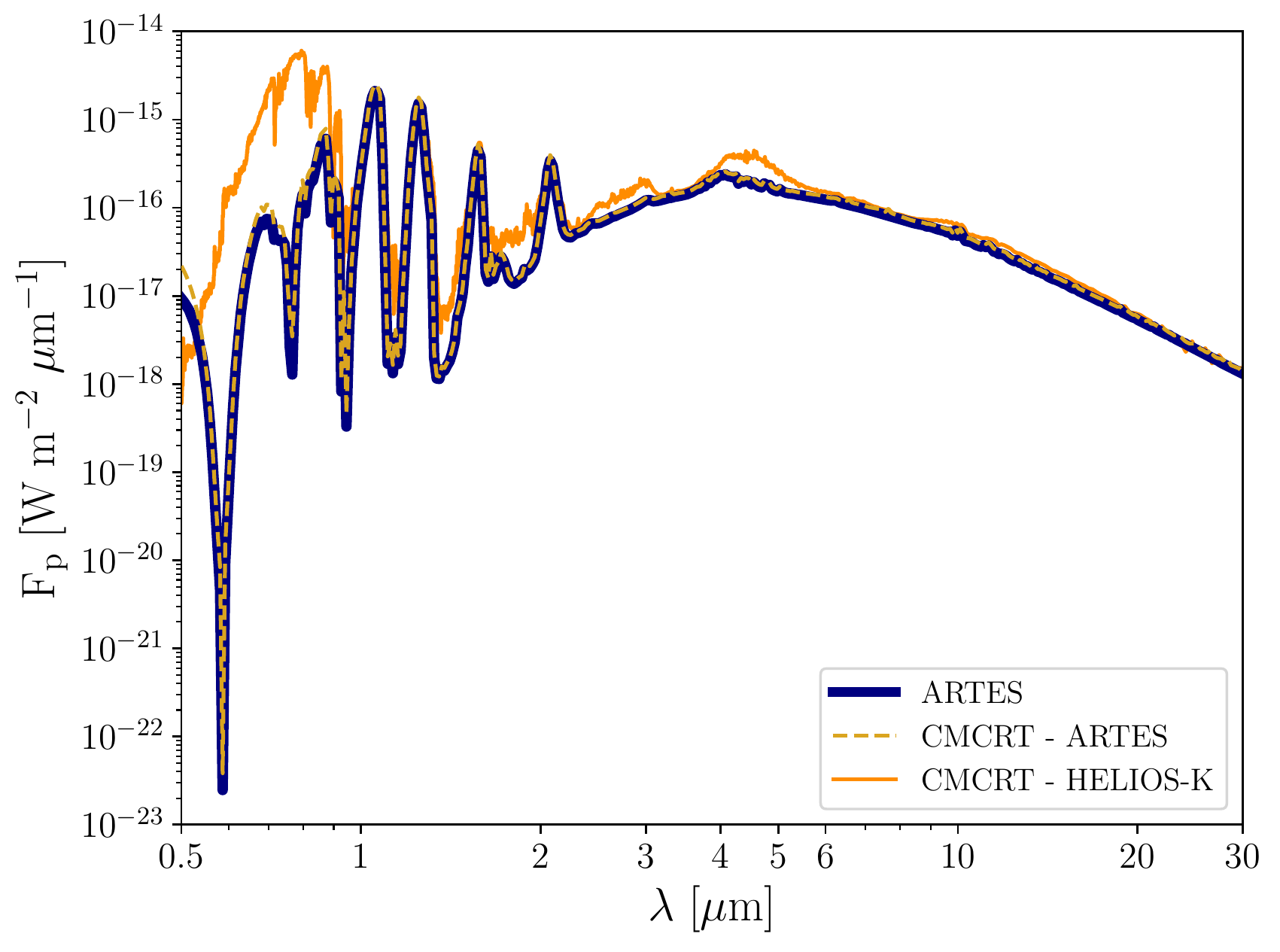}
   \includegraphics[width=0.48\textwidth]{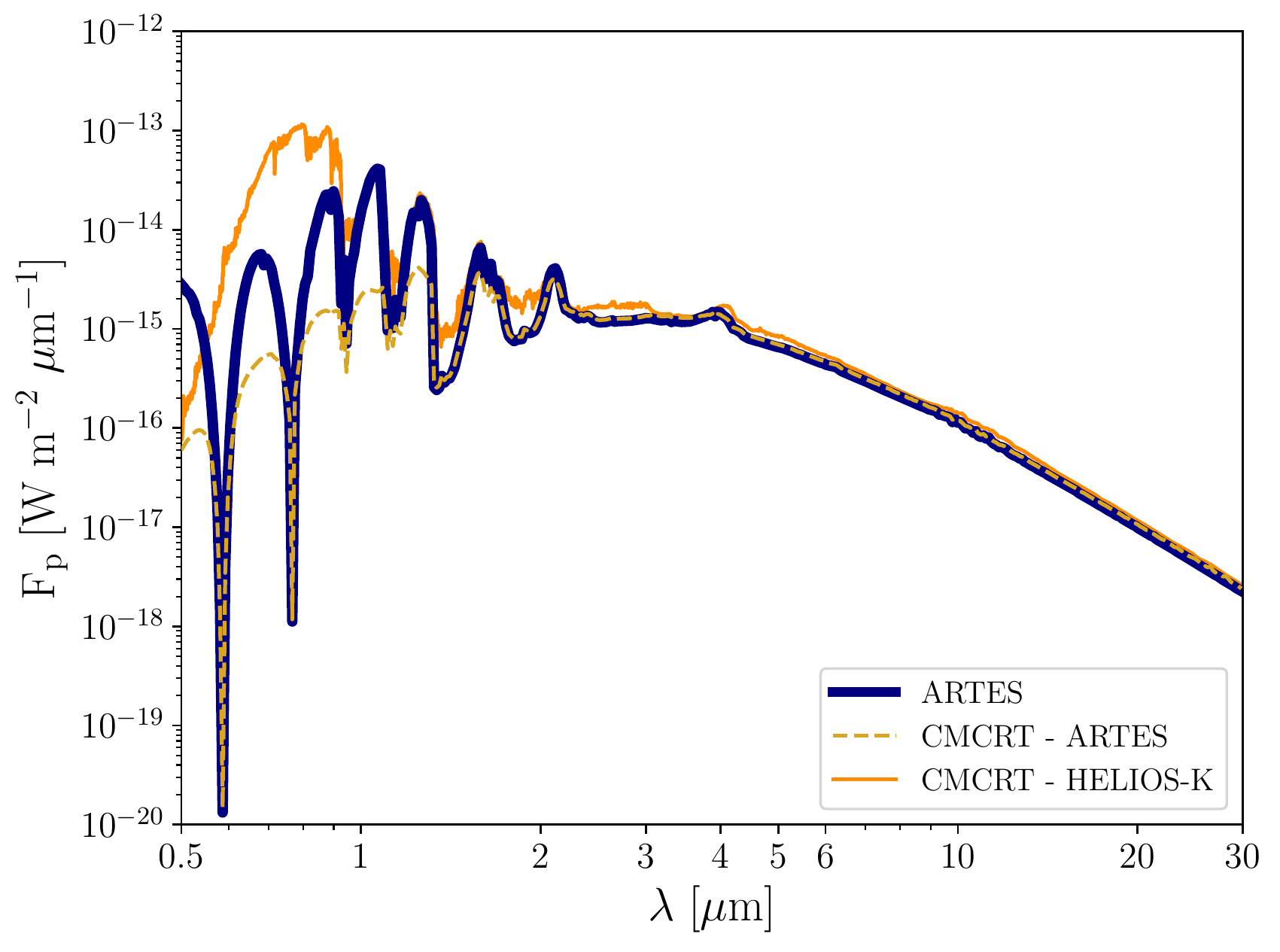}
   \caption{Comparison to the emission spectra benchmarks performed in \citet{Stolker2017}.
   Original \citet{Stolker2017} data (dark blue, solid line), CMCRT using the \citet{Stolker2017} opacities (gold, dashed line) and
   CMCRT using the HELIOS-K opacities (orange, solid line).
   Top: T$_{\rm eff}$ = 400 K benchmark.
   Middle: T$_{\rm eff}$ = 800 K benchmark.
   Bottom: T$_{\rm eff}$ = 1200 K benchmark.}
   \label{fig:Stolker_res}
\end{figure}

In \citet{Stolker2017} several emission spectra benchmarks were performed using the 3D MCRT model ARTES, primarily to investigate emergent polarisation signatures from directly imaged exoplanets.
We repeat the cloud free, self-luminous planet emission spectra protocol from Appendix B in \citet{Stolker2017}, denoted by T$_{\rm eff}$ = 400 K, 800 K and 1200 K respectively (Fig. \ref{fig:TP}).
We run two simulations for each profile, one with the mean opacity tables directly used in \citet{Stolker2017} (T. Stolker priv. corr.) in individual wavelength mode and one using the HELIOS-K opacities in correlated-k mode.

Figure \ref{fig:Stolker_res} presents a comparison between our results and \citet{Stolker2017}.
The simulations using the \citet{Stolker2017} opacity tables are in excellent agreement with the ARTES output, except at optical wavelengths in the T$_{\rm eff}$ = 1200 K test case.
The results using the HELIOS-k opacities show differences in the optical regime due to the non-inclusion of Na and K opacity, but also show a general positive offset with shallower absorption features in the infrared.
We suggest that the differences in the results produced by CMCRT when using the \citet{Stolker2017} opacity tables and HELIOS-K tables stem from the choice of Voigt line wing cutoff (Sect. \ref{sec:discussion}) used for the input opacities between the models.
However, it is encouraging that CMCRT produces consistent results to the \citet{Stolker2017} spectra when using the \textsc{ARTES} opacities directly, as these two models share similar 3D Monte Carlo methodologies.

\subsection{Baudino et al. Benchmarks}
\label{sec:Baudino}

\begin{figure} 
   \centering
   \includegraphics[width=0.48\textwidth]{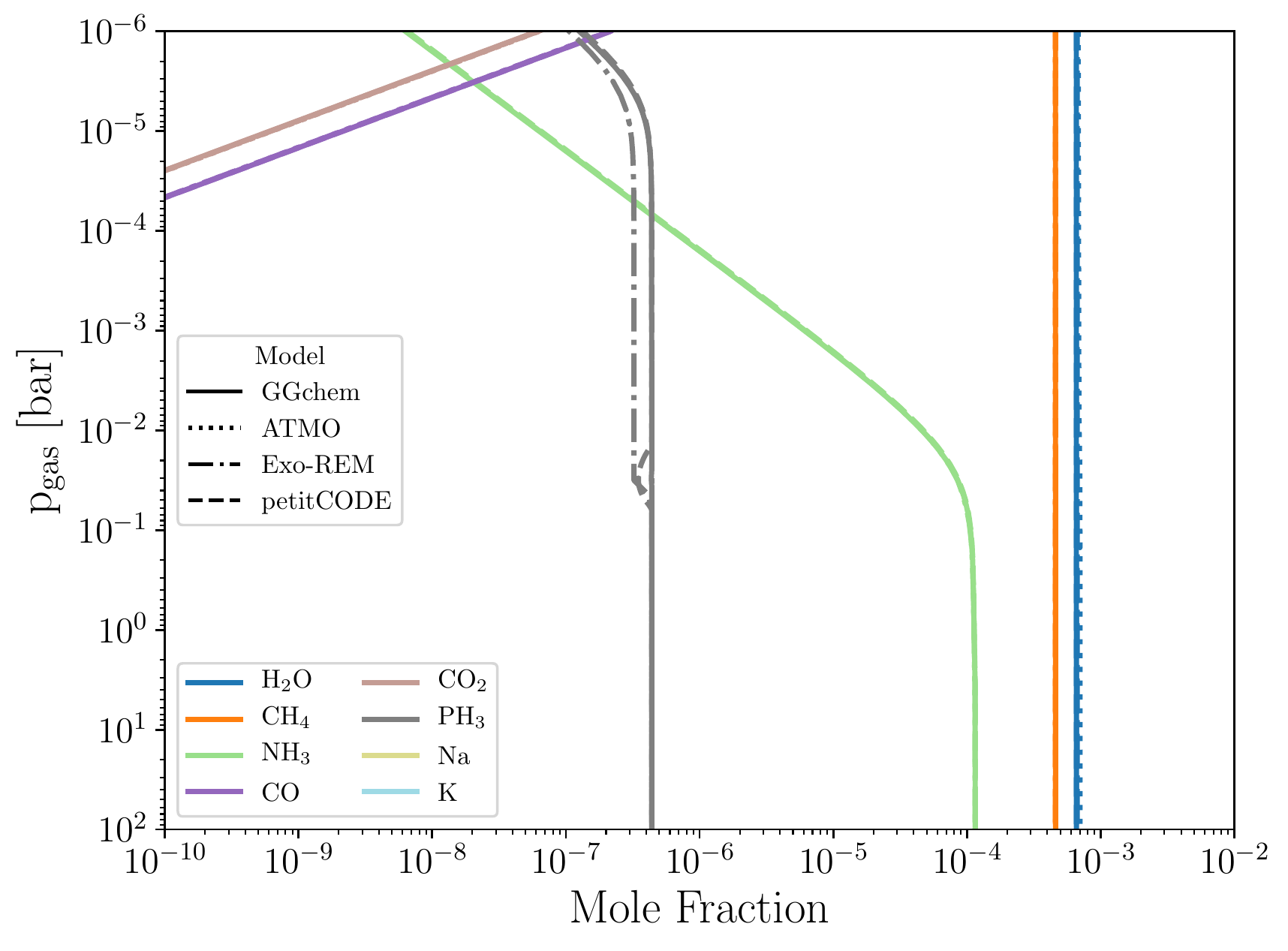}
   \includegraphics[width=0.48\textwidth]{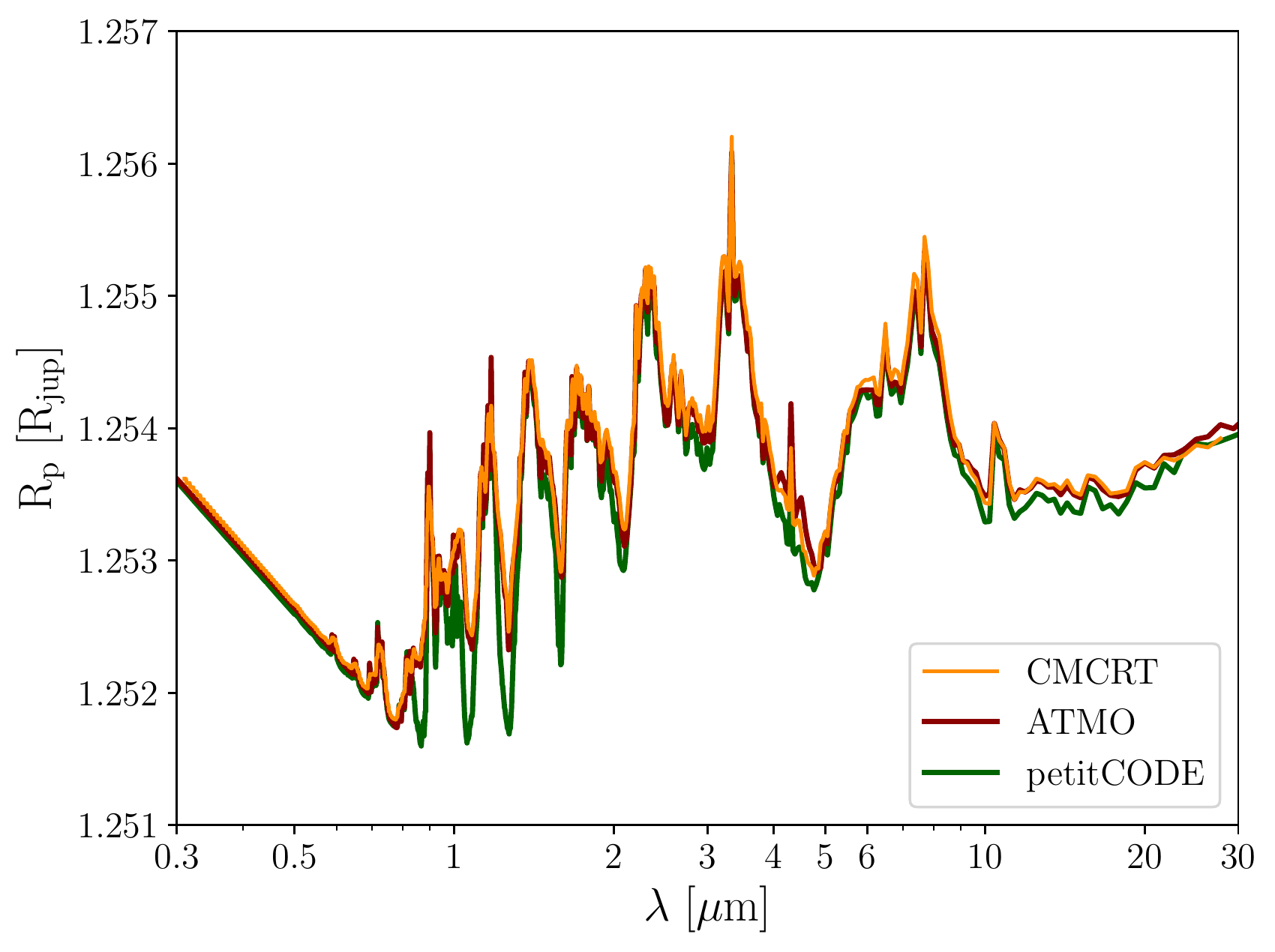}
   \includegraphics[width=0.48\textwidth]{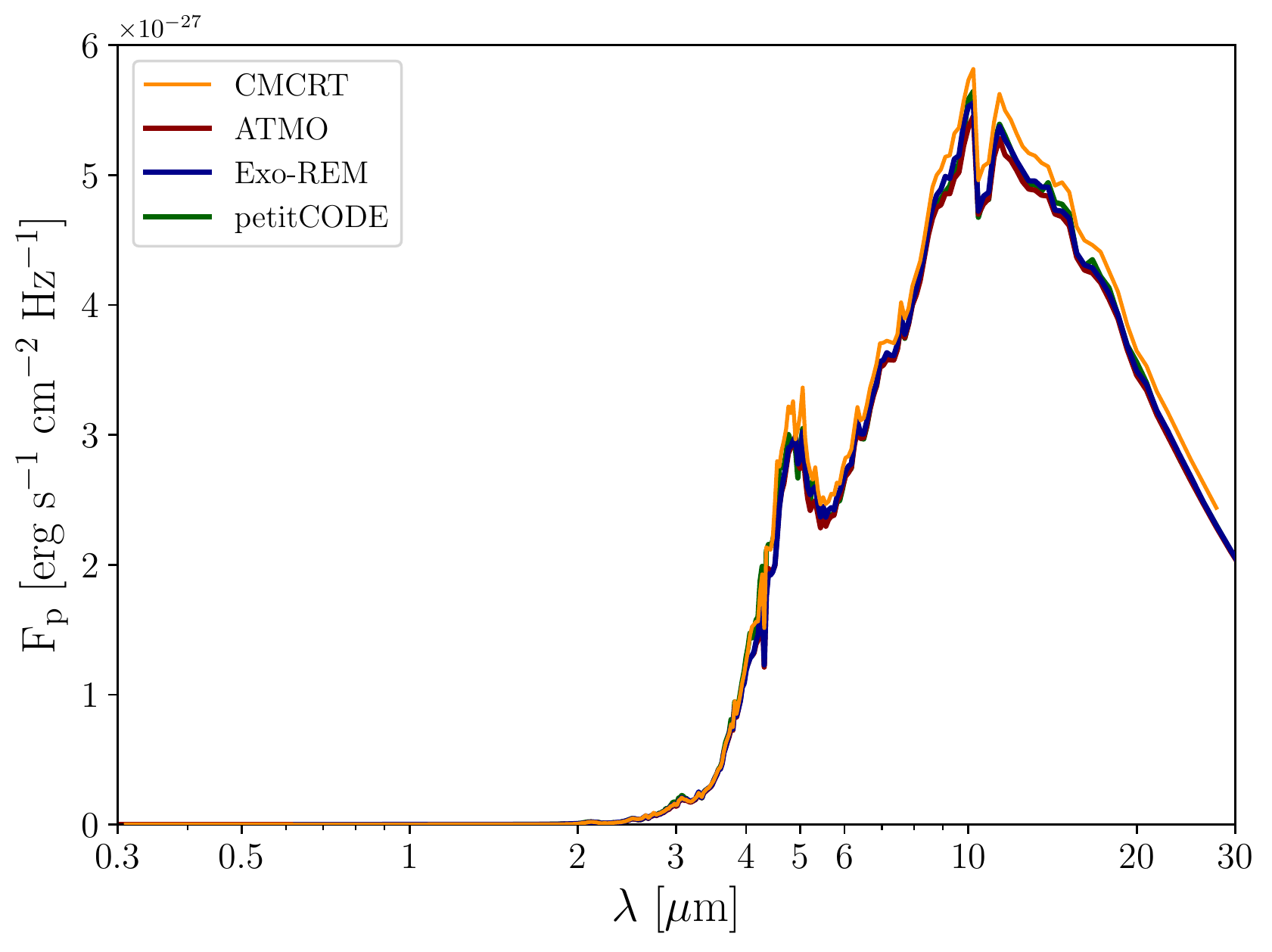}
   \caption{Benchmark results for the $T_{\rm eff}$ = 500 K prescribed T-p profile from \citet{Baudino2017}.
    Top: Mole ratio results; solid lines show the results of the GGchem code, used in this study.
    Dotted, dot-dash and dashed lines denote the results from ATMO, Exo-REM and petitCODE respectively from \citet{Baudino2017}.
    Middle: Transit spectra results.
    Bottom: Emission spectra results.
    Solid orange lines show the results of the MCRT transmission (Sect. \ref{sec:trans_spec}) and emission (Sect. \ref{sec:em_spec}) method used in this study.
    Solid red, blue and green lines represent the ATMO, Exo-REM and petitCODE results respectively from \citet{Baudino2017}. }
   \label{fig:500K_res}
\end{figure}

\begin{figure} 
   \centering
   \includegraphics[width=0.48\textwidth]{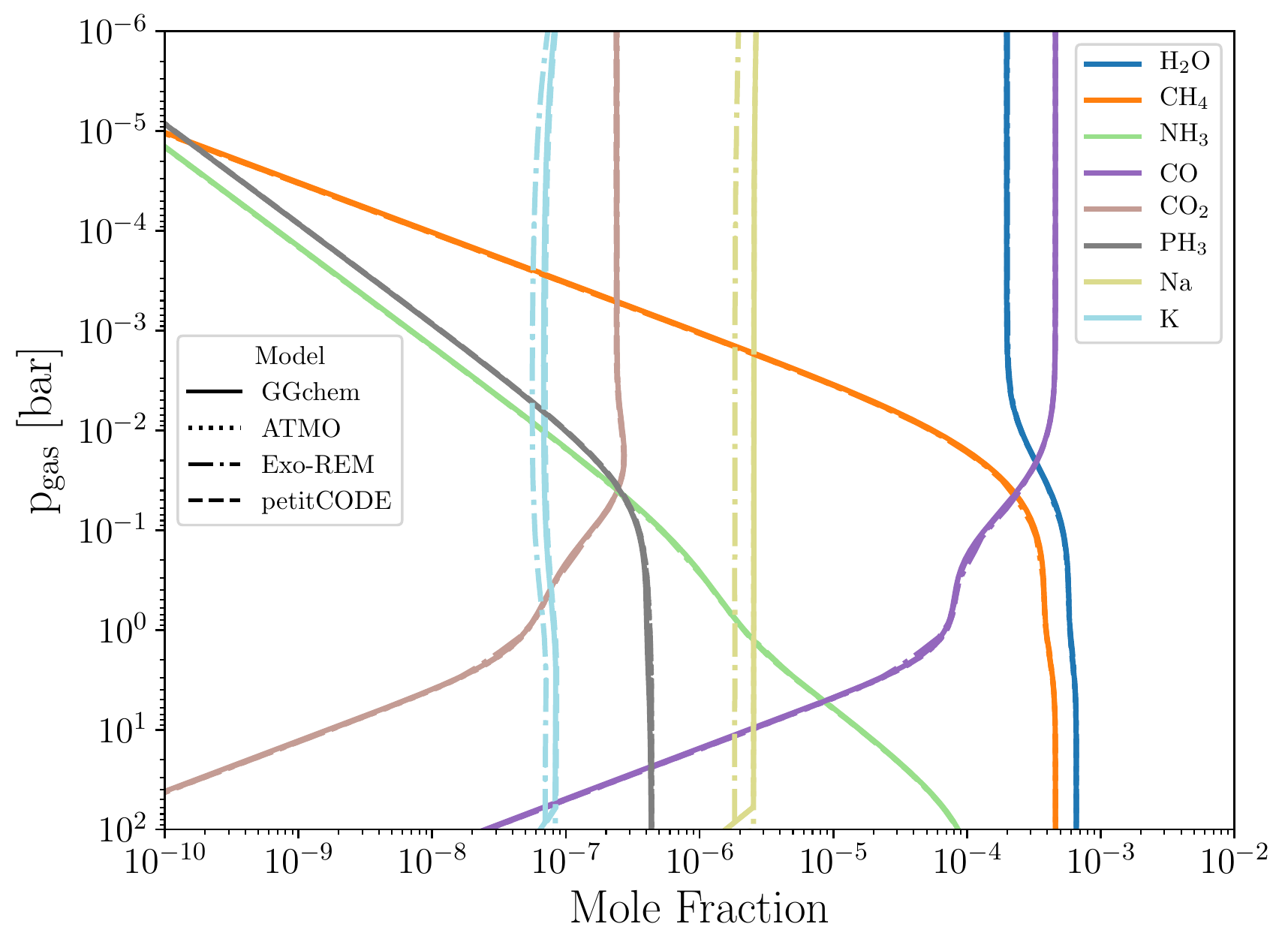}
   \includegraphics[width=0.48\textwidth]{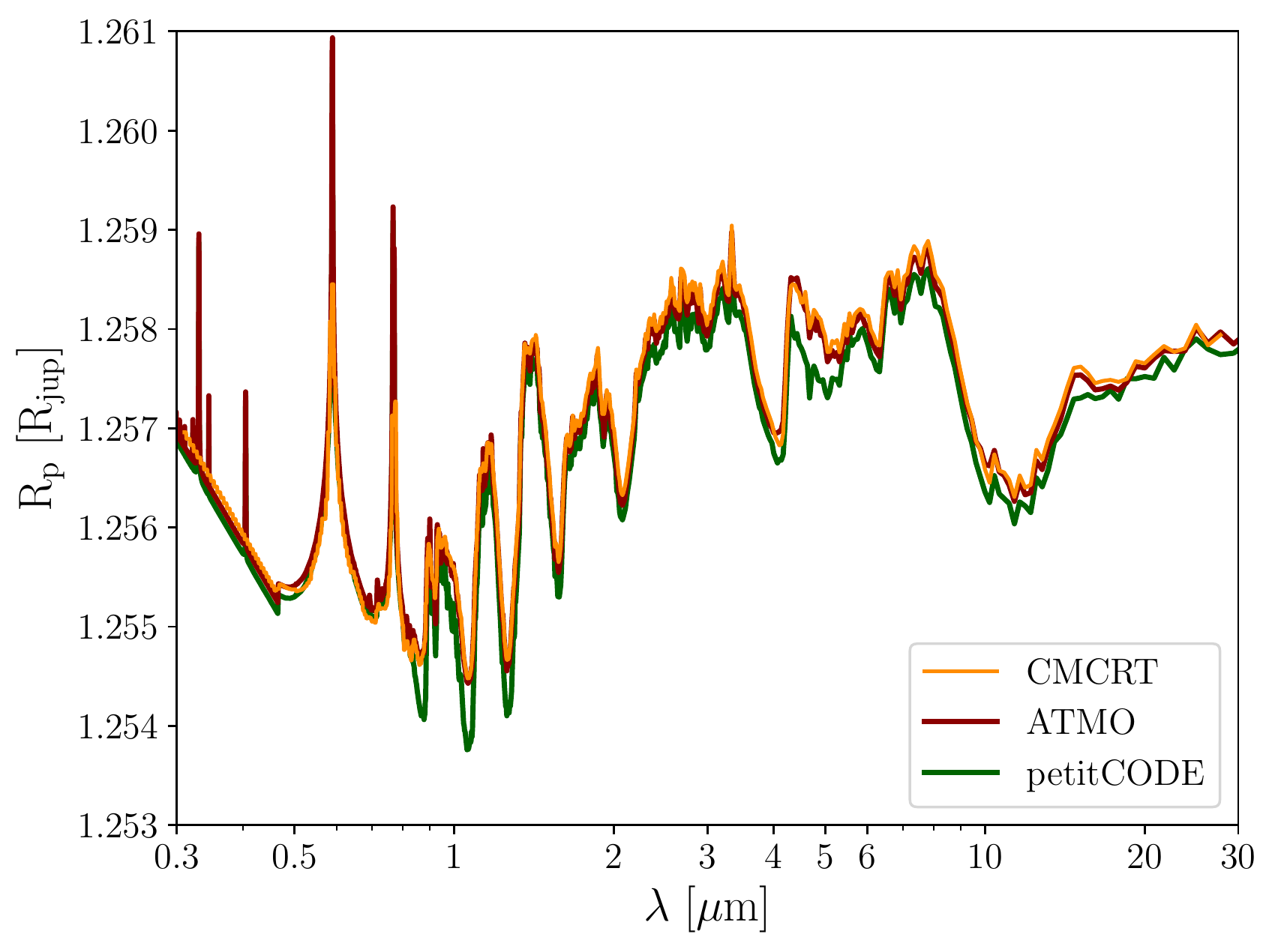}
   \includegraphics[width=0.48\textwidth]{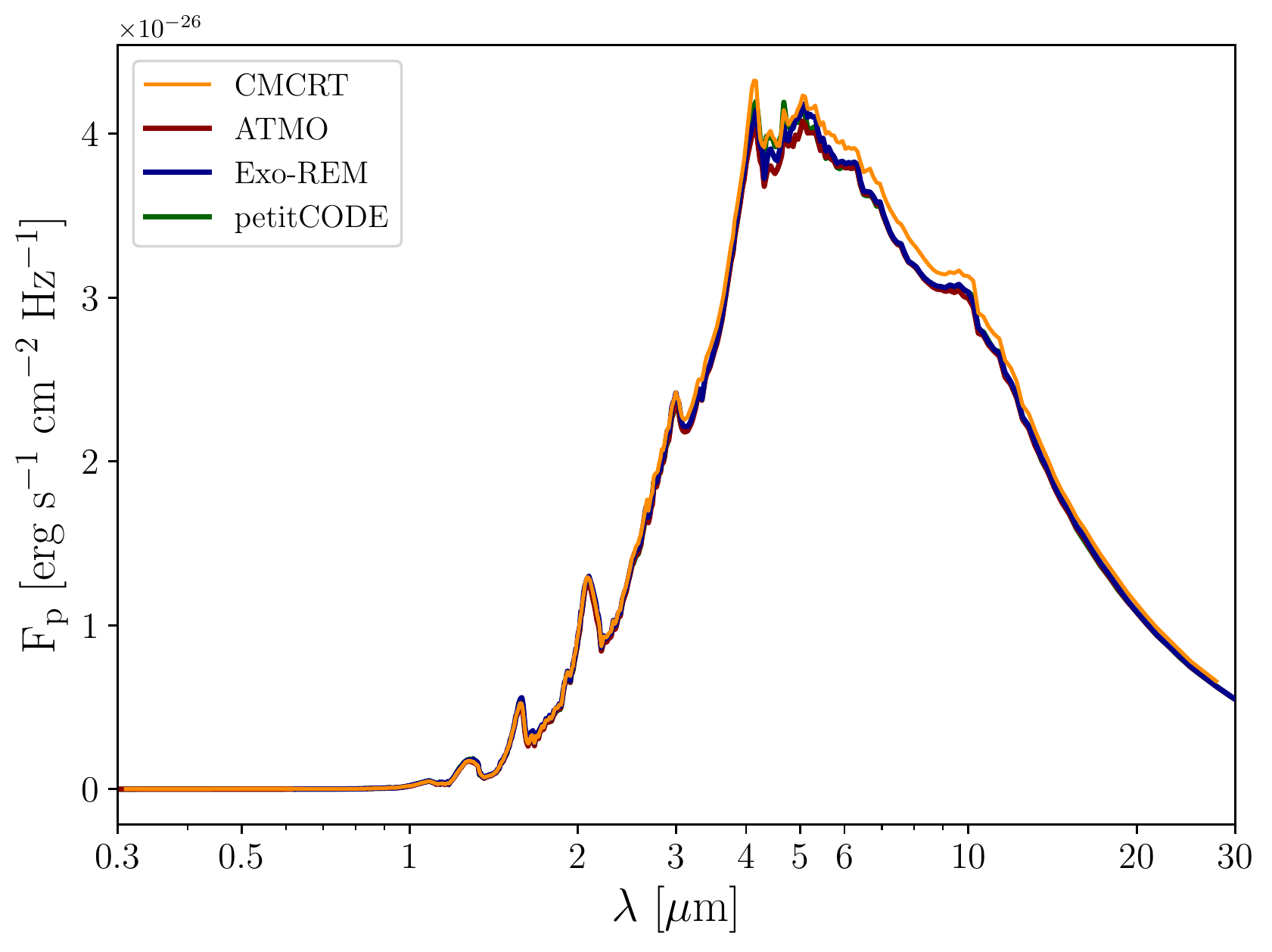}
   \caption{Benchmark results for the $T_{\rm eff}$ = 1000 K prescribed T-p profile from \citet{Baudino2017}.
    See the Fig. \ref{fig:500K_res} caption for a detailed description.}
   \label{fig:1000K_res}
\end{figure}

\begin{figure} 
   \centering
   \includegraphics[width=0.48\textwidth]{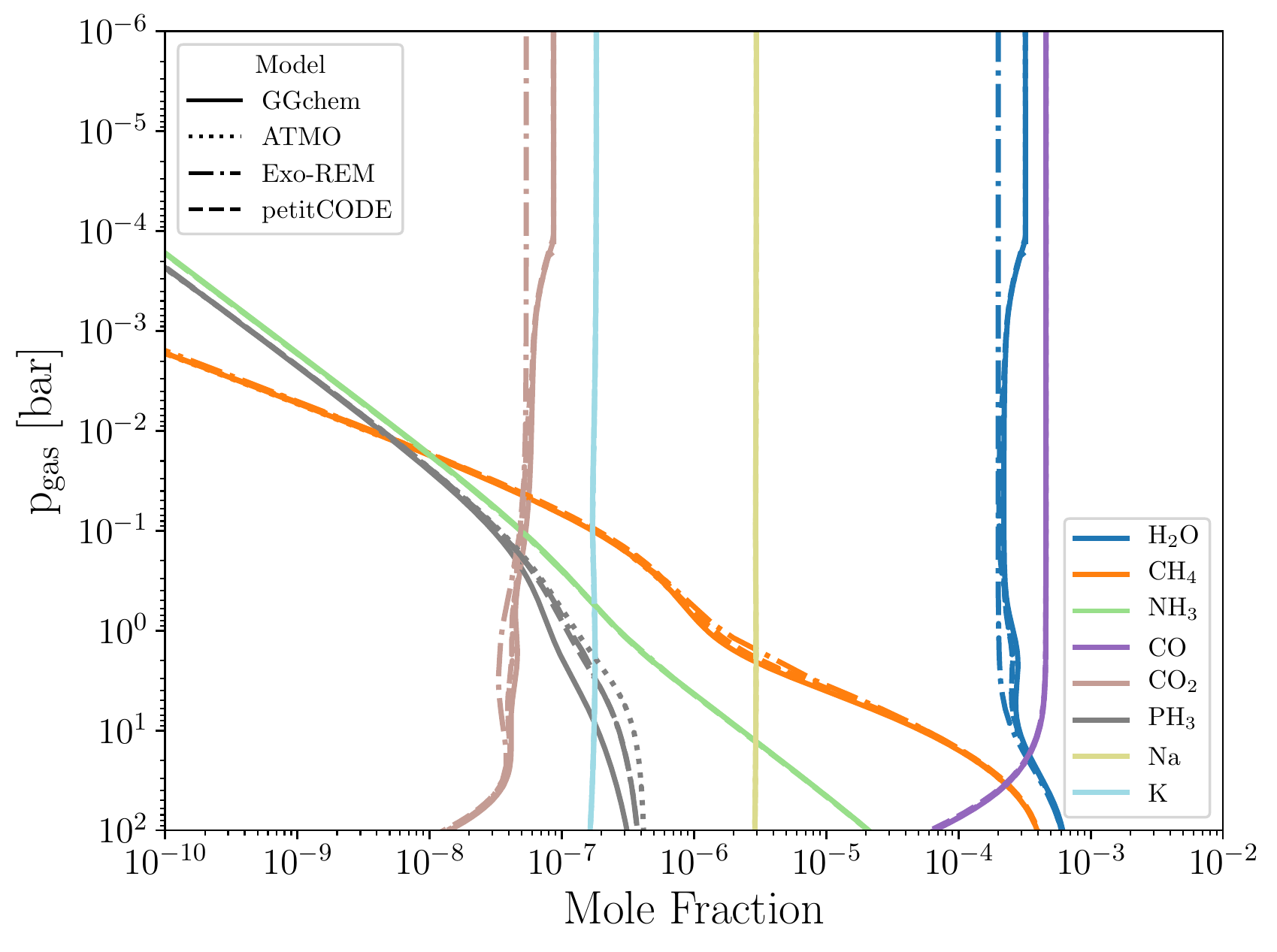}
   \includegraphics[width=0.48\textwidth]{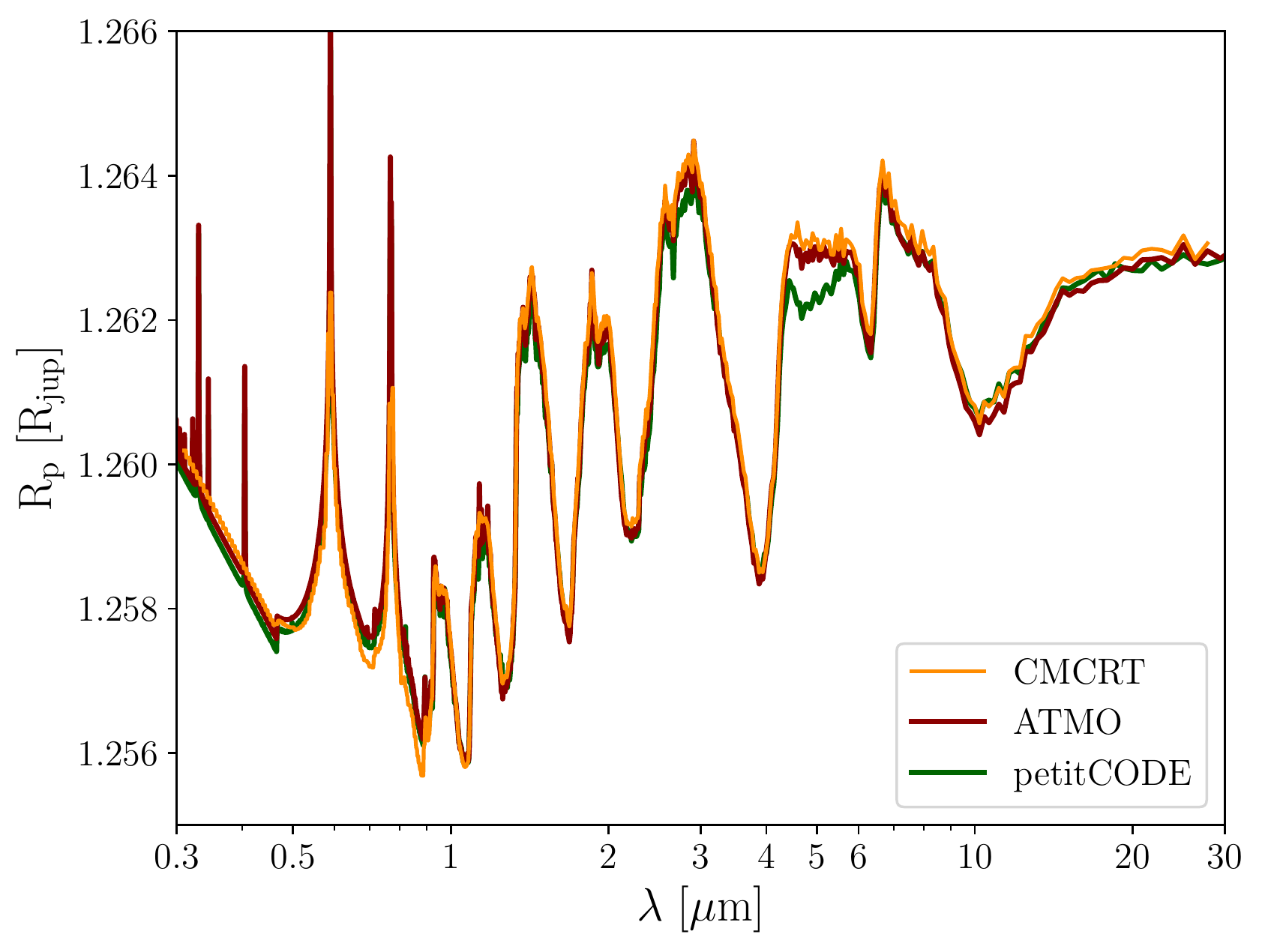}
   \includegraphics[width=0.48\textwidth]{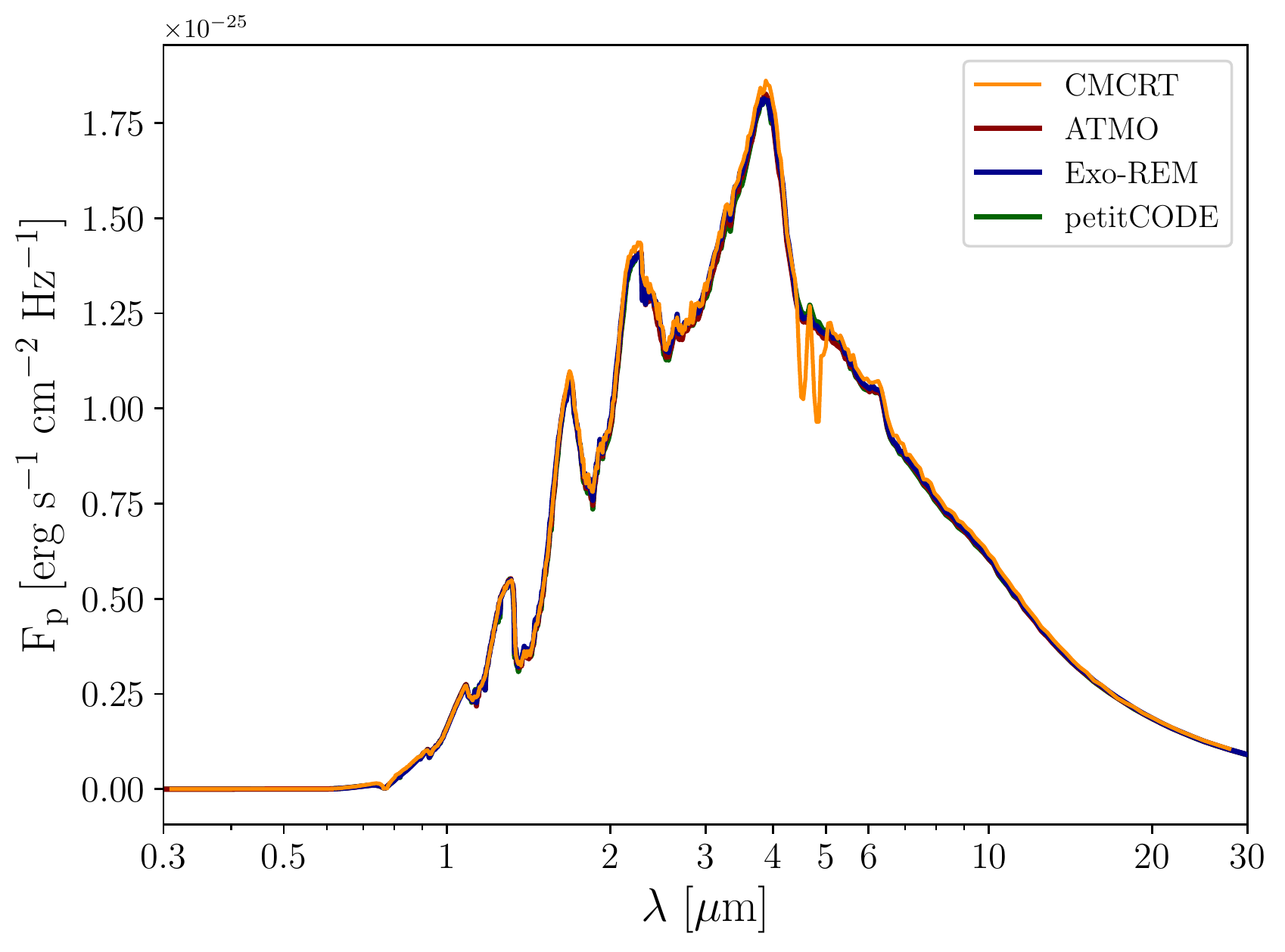}
   \caption{Benchmark results for the $T_{\rm eff}$ = 1500 K prescribed T-p profile from \citet{Baudino2017}.
  See the Fig. \ref{fig:500K_res} caption for a detailed description. }
   \label{fig:1500K_res}
\end{figure}

\begin{figure} 
   \centering
   \includegraphics[width=0.48\textwidth]{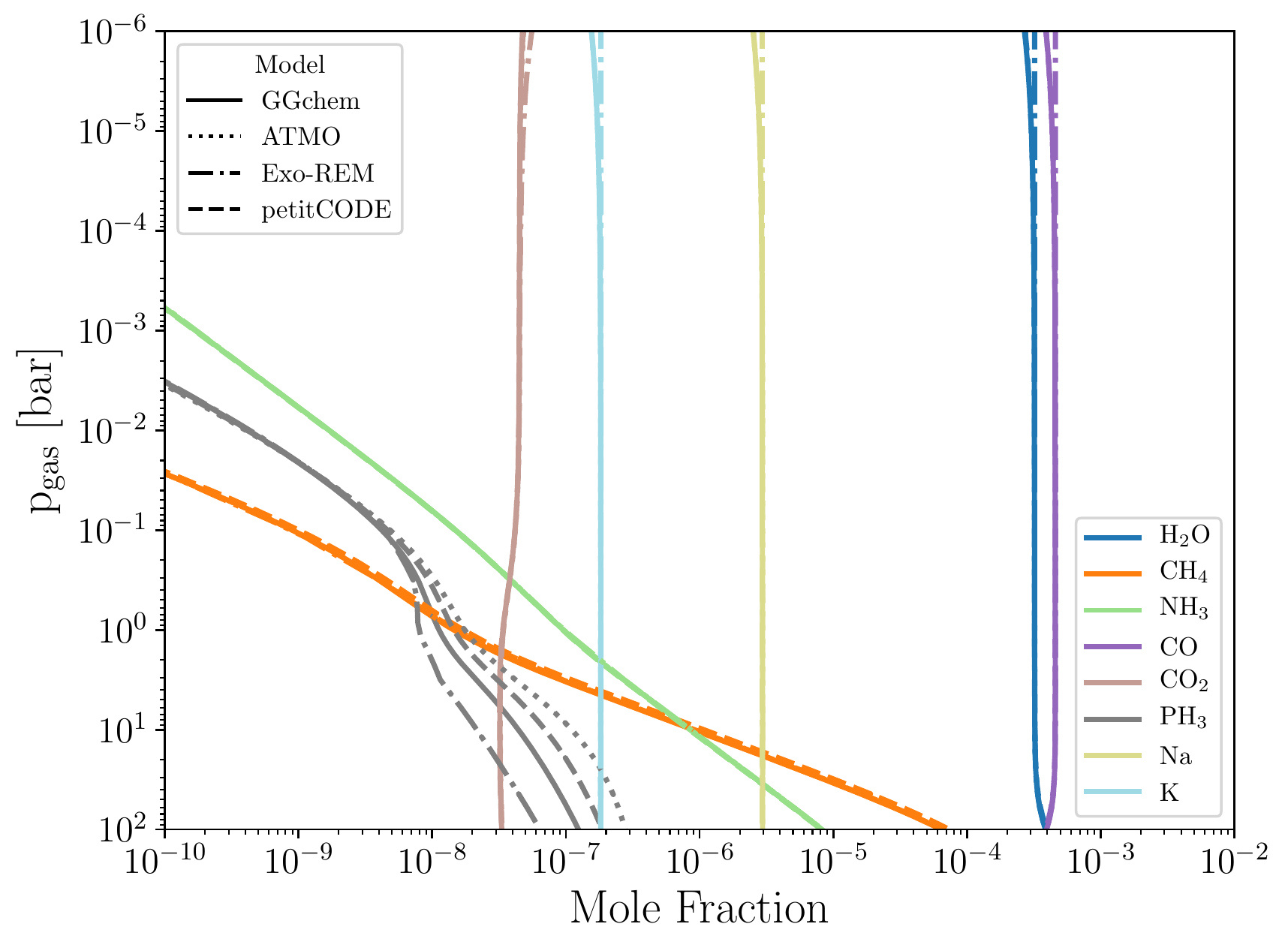}
   \includegraphics[width=0.48\textwidth]{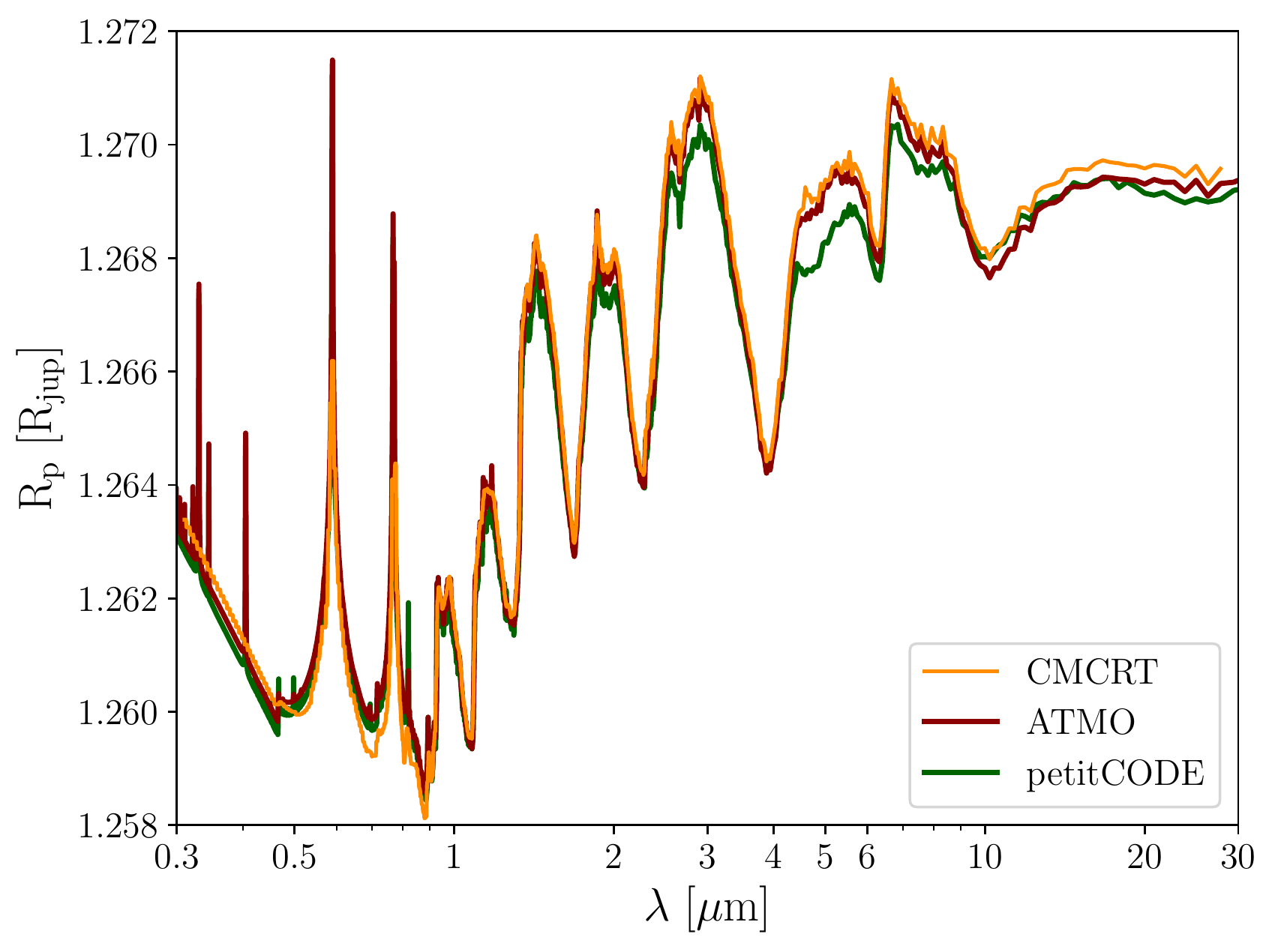}
   \includegraphics[width=0.48\textwidth]{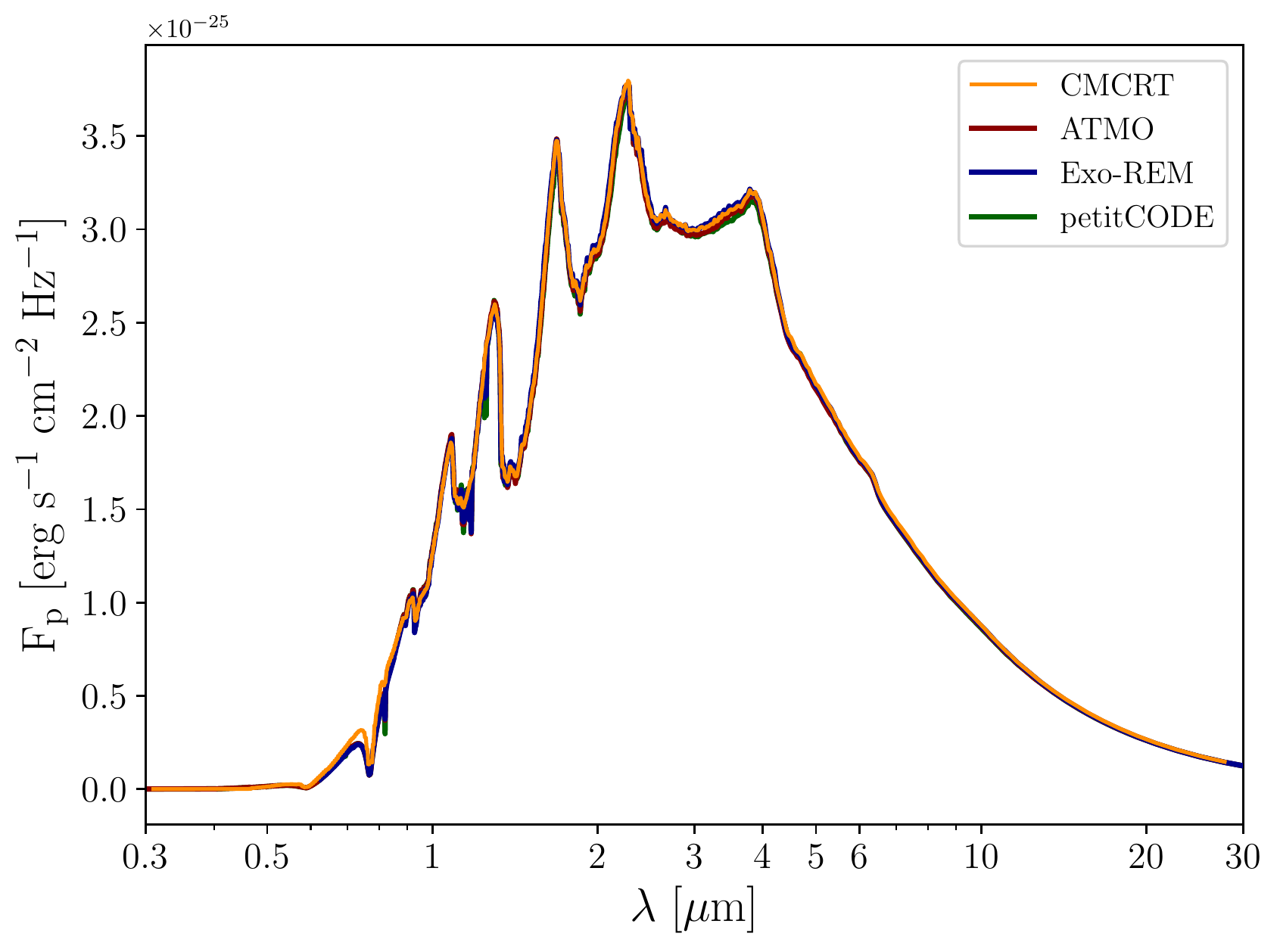}
   \caption{Benchmark results for the $T_{\rm eff}$ = 2000 K prescribed T-p profile from \citet{Baudino2017}.
  See the Fig. \ref{fig:500K_res} caption for a detailed description.}
   \label{fig:2000K_res}
\end{figure}

\begin{figure} 
   \centering
   \includegraphics[width=0.48\textwidth]{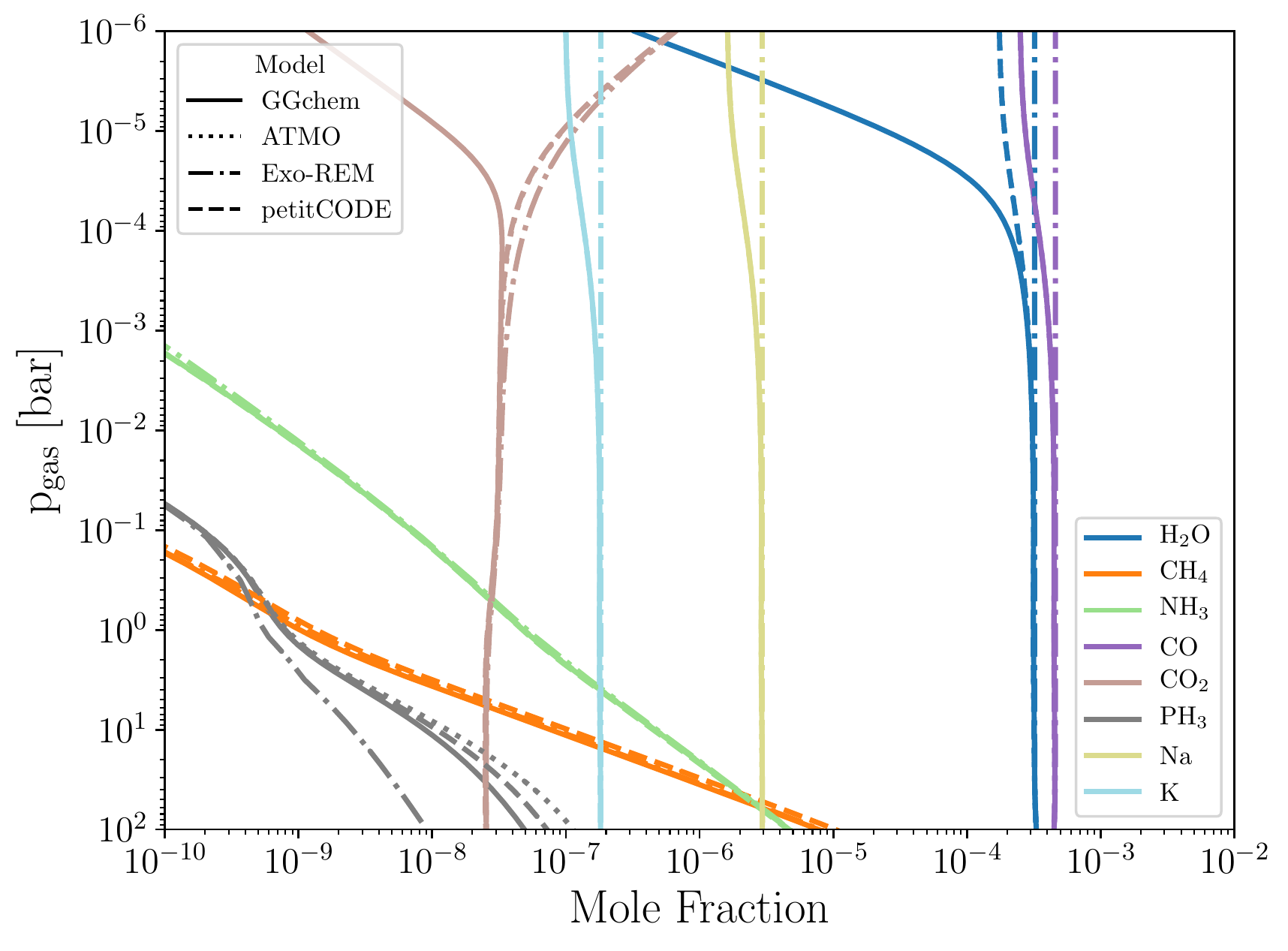}
   \includegraphics[width=0.48\textwidth]{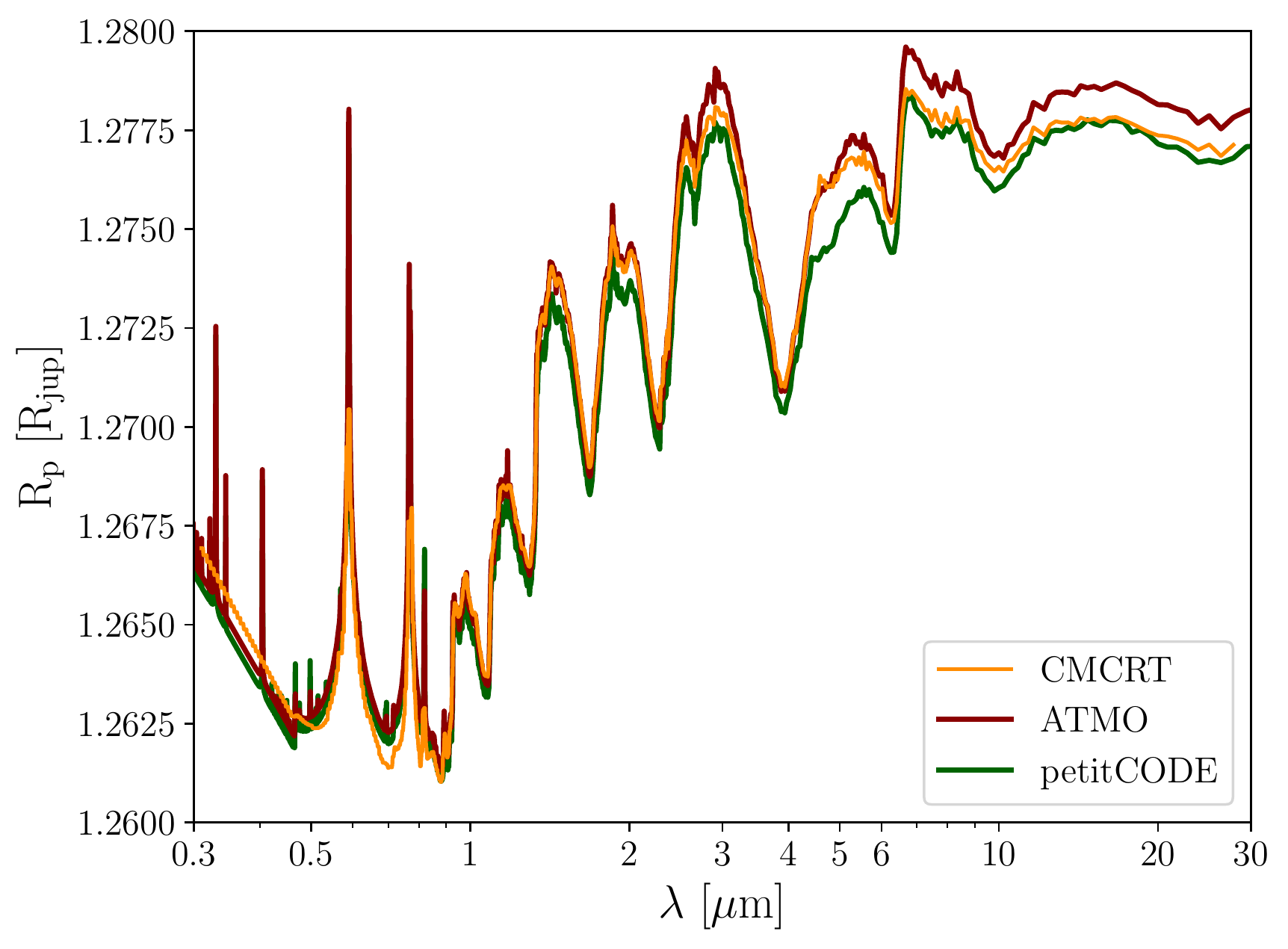}
   \includegraphics[width=0.48\textwidth]{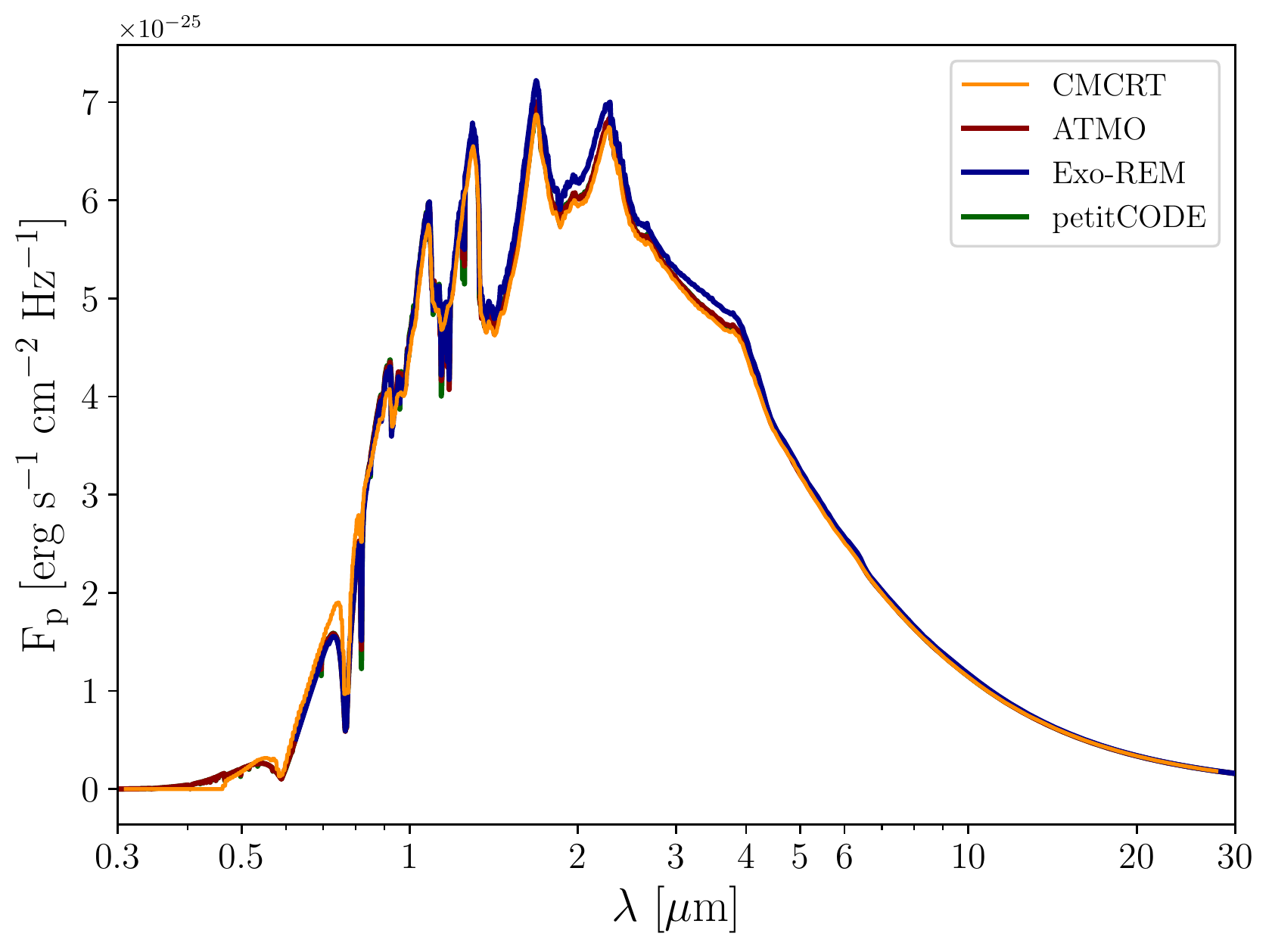}
   \caption{Benchmark results for the $T_{\rm eff}$ = 2500 K prescribed T-p profile from \citet{Baudino2017}.
  See the Fig. \ref{fig:500K_res} caption for a detailed description.}
   \label{fig:2500K_res}
\end{figure}

In \citet{Baudino2017}, the radiative-convective models \textsc{ATMO} \citep{Tremblin2015, Drummond2016, Goyal2018}, \textsc{Exo-REM} \citep{Baudino2015} and \textsc{petitCODE} \citep{Molliere2015,Molliere2017} were benchmarked across a variety of 1D temperature and pressure conditions suitable for exoplanet atmospheres.
We benchmark to the transmission and emission spectra of the prescribed \citet{Guillot2010} T-p profiles, denoted as T$_{\rm eff}$ = 500 K, 1000 K, 1500 K, 2000 K and 2500 K (Fig. \ref{fig:TP}).
Figures \ref{fig:500K_res}, \ref{fig:1000K_res}, \ref{fig:1500K_res}, \ref{fig:2000K_res} and \ref{fig:2500K_res} present the T$_{\rm eff}$ = 500 K, 1000 K, 1500 K, 2000 K and 2500 K benchmark results respectively.
For these tests we apply the NEMESIS k-distribution tables as the input opacities.
All transmission and emission results from CMCRT are convolved to the benchmark wavelength resolution using the \textsc{SpectRes} package \citep{Carnall2017}.

\subsubsection{Molecular abundances}

In \citet{Baudino2017} the chemical equilibrium (CE) schemes of each model were compared and used as input gas phase abundances.
We apply the publicly available CE with condensation code of \citet{Woitke2018}, \textsc{GGchem}, to each T-p profile and reduce the full species database \citep{Worters2017} to the gas and solid/liquid phase species listed in \citet{Baudino2017}.
Input elemental ratios at solar metallicity are taken from \citet{Asplund2009}.
As per the benchmark protocols, H$_{3}$PO$_{4}$[l] (Phosphoric acid) is added to the \textsc{GGchem} condensate list, and we additionally include the calculation of gas phase SiO since the vapour pressure expression from \citet{Wetzel2013} is used to calculate the SiO[s] supersaturation ratio.
The corrected thermochemical data for PH$_{3}$ from \citet{Lodders1999b} is also used.

The CE calculations compare well to the other codes, with differences occurring for PH$_{3}$ mole fractions at high pressures in the $T_{\rm eff}$ = 1500 K, 2000 K and 2500 K profiles.
A significant deviation between the calculations is seen in the $T_{\rm eff}$ = 2000 K and 2500 K profiles for $p_{\rm gas}$ $<$ 10$^{-3}$ bar where the H$_{2}$O, CO, CO$_{2}$, Na and K abundances drop off.
We attribute this to the thermal disassociation and ionisation of these molecular and atomic species at lower pressure and higher ($T_{\rm gas}$ $>$ 2000 K) gas temperature atmospheric conditions.
We note that considering thermal dissociation of molecules was not explicitly included in the \citet{Baudino2017} benchmark protocols.

\subsubsection{Transmission spectra}

The transmission spectra for each profile is typically consistent to the benchmark spectra to within the $\approx$ 10s of ppm level.
CMCRT compares best with the ATMO results, but are generally slightly positively offset from the ATMO results.
We suggest a 3D grid effect in our methodology is responsible for this positive offset (Sect. \ref{sec:discussion}).

\subsubsection{Emission spectra}

The emission spectra are generally consistent with the benchmark results to within the $\approx$ 10\% level.
A small positive offset between CMCRT and the benchmark results is seen in the T$_{\rm eff}$ = 500 K and 1000 K profile at the peak emission and the Rayleigh-Jeans tail wavelengths.
We suggest a difference between the 3D spherical and plane-parallel models produces this offset (Sect. \ref{sec:discussion}).
As in Sect. \ref{sec:NEMESIS}, for the T$_{\rm eff}$ = 1500 K profile a CO absorption feature is present in the CMCRT model between 4-5 $\mu$m not seen in the benchmark results.
Since all the models in this test used the \citet{Rothman2010} CO line-list, this suggests that the CO line forming regions in the deep atmosphere are different in the 3D grid compared to the 1D codes for this profile.

\section{Post-processing GCM output}
\label{sec:GCM}

\begin{figure*} 
   \centering
   \includegraphics[width=0.48\textwidth]{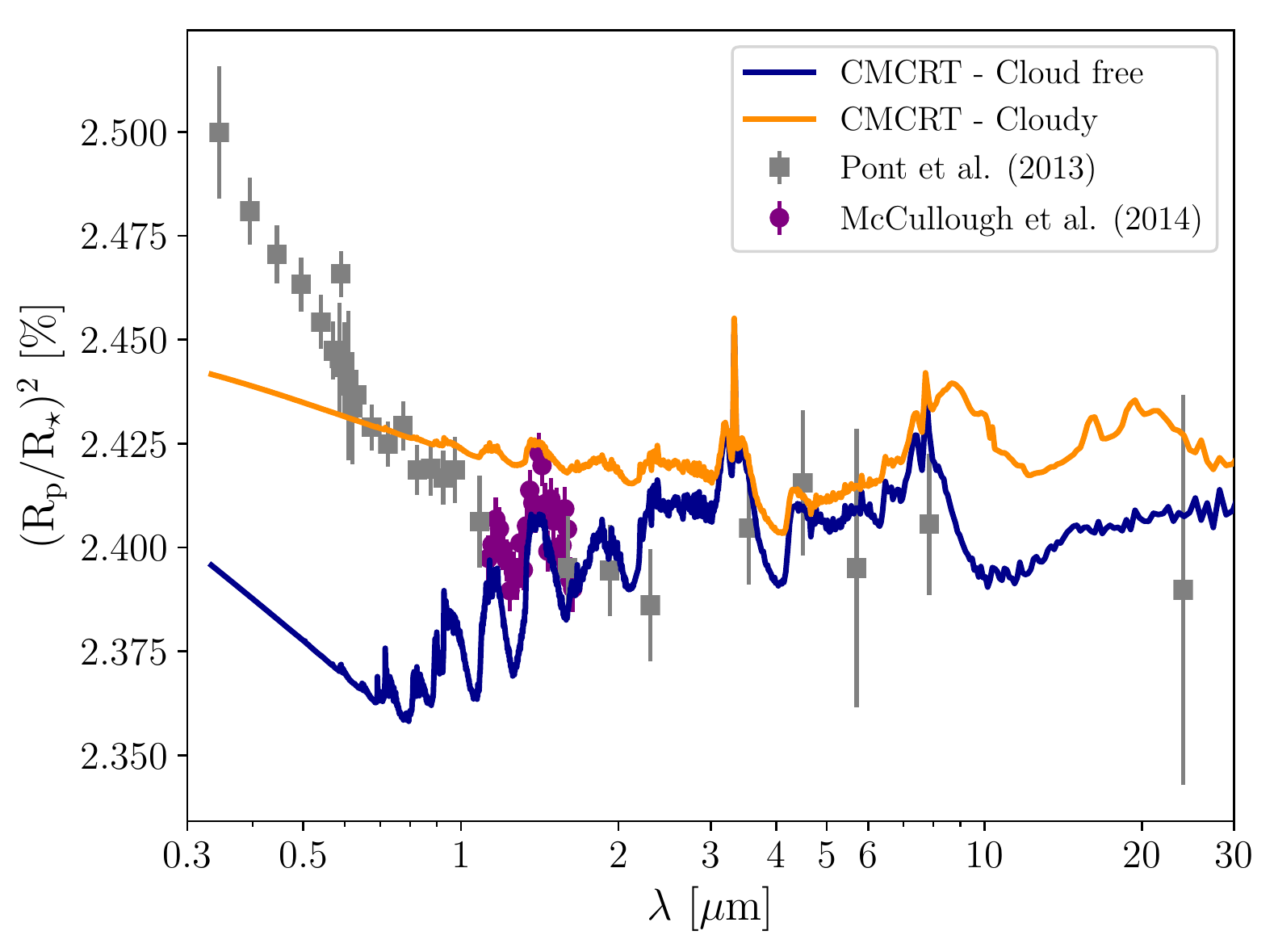}
   \includegraphics[width=0.48\textwidth]{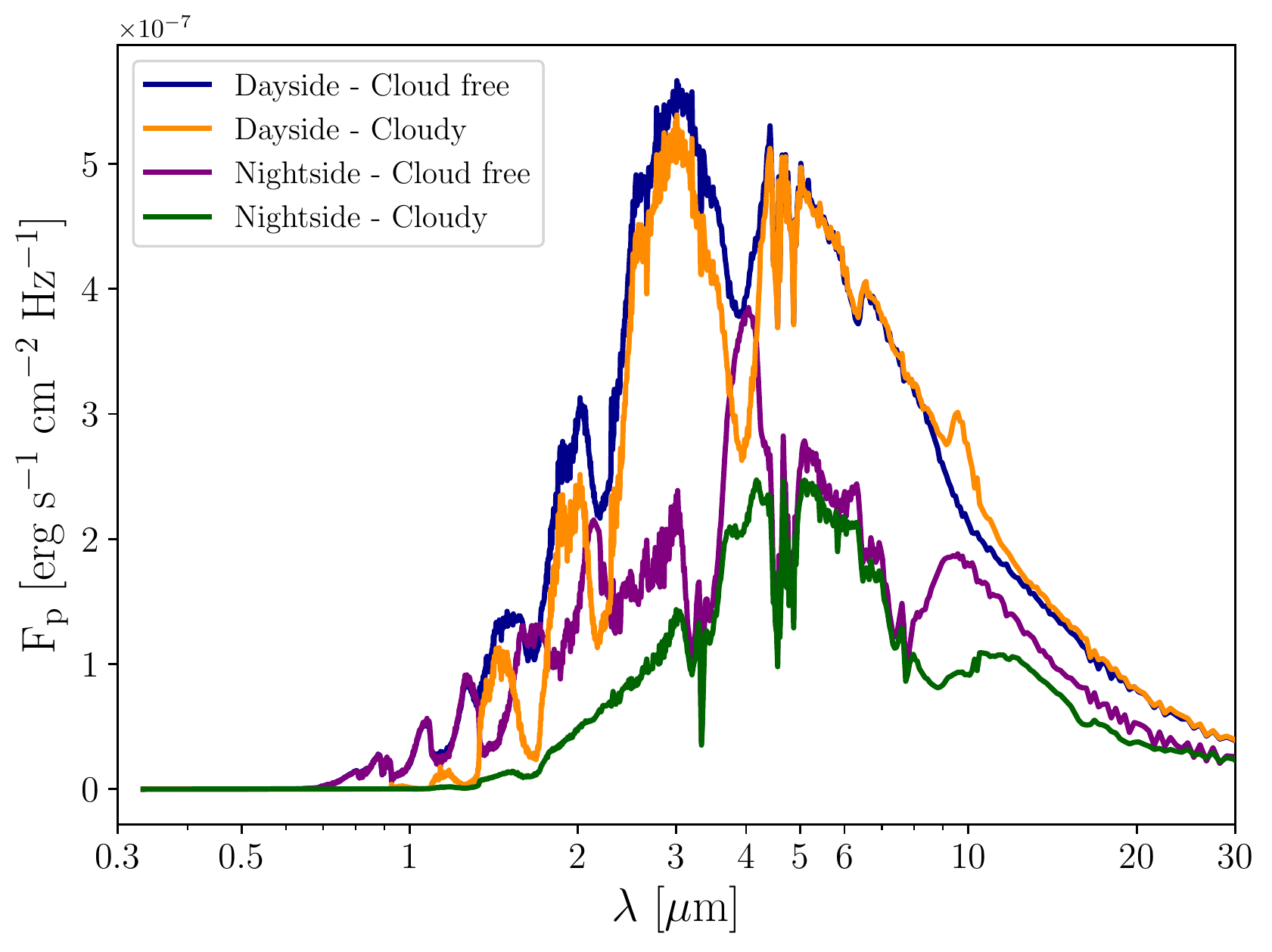}
   \includegraphics[width=0.48\textwidth]{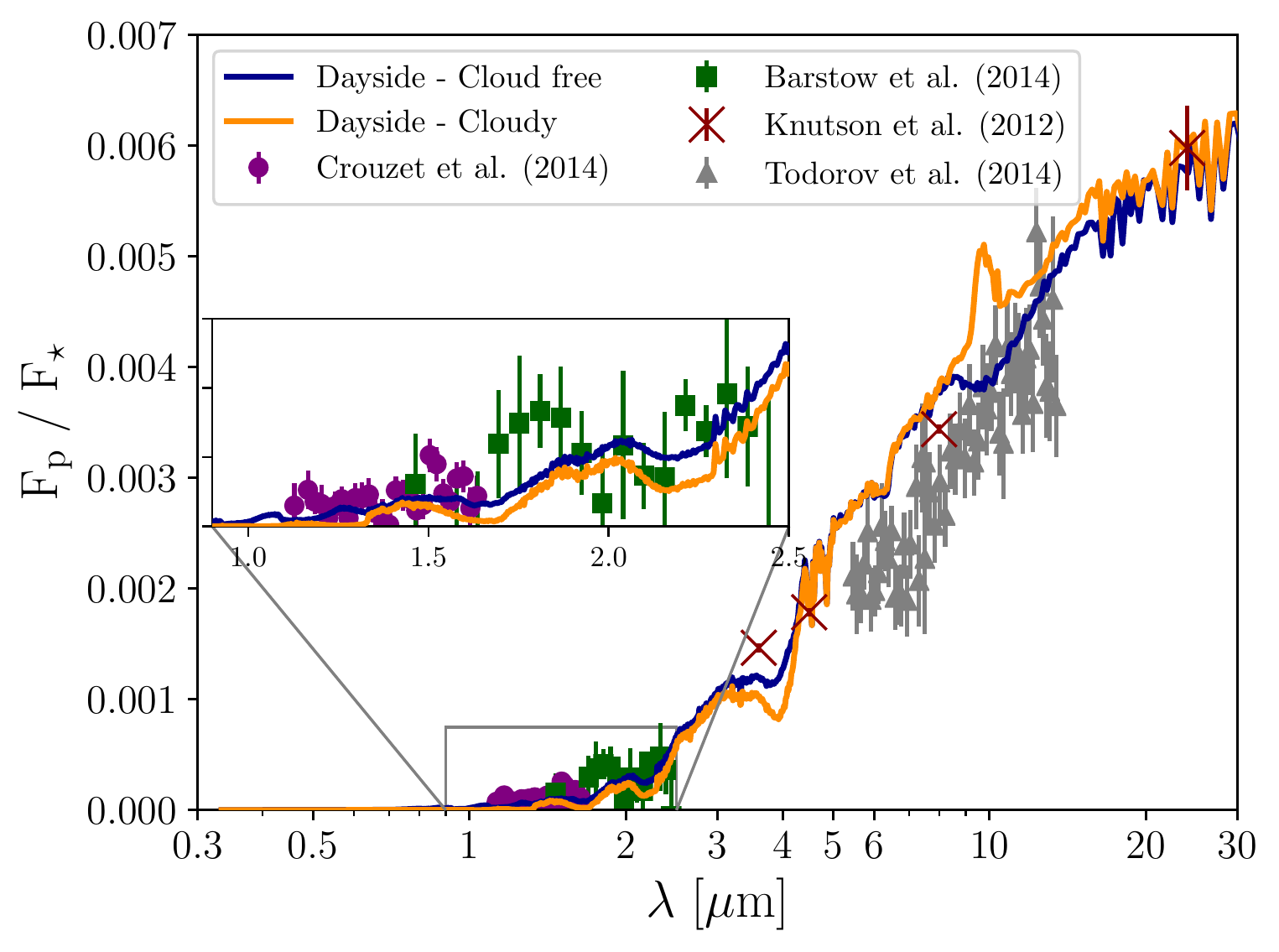}
   \includegraphics[width=0.48\textwidth]{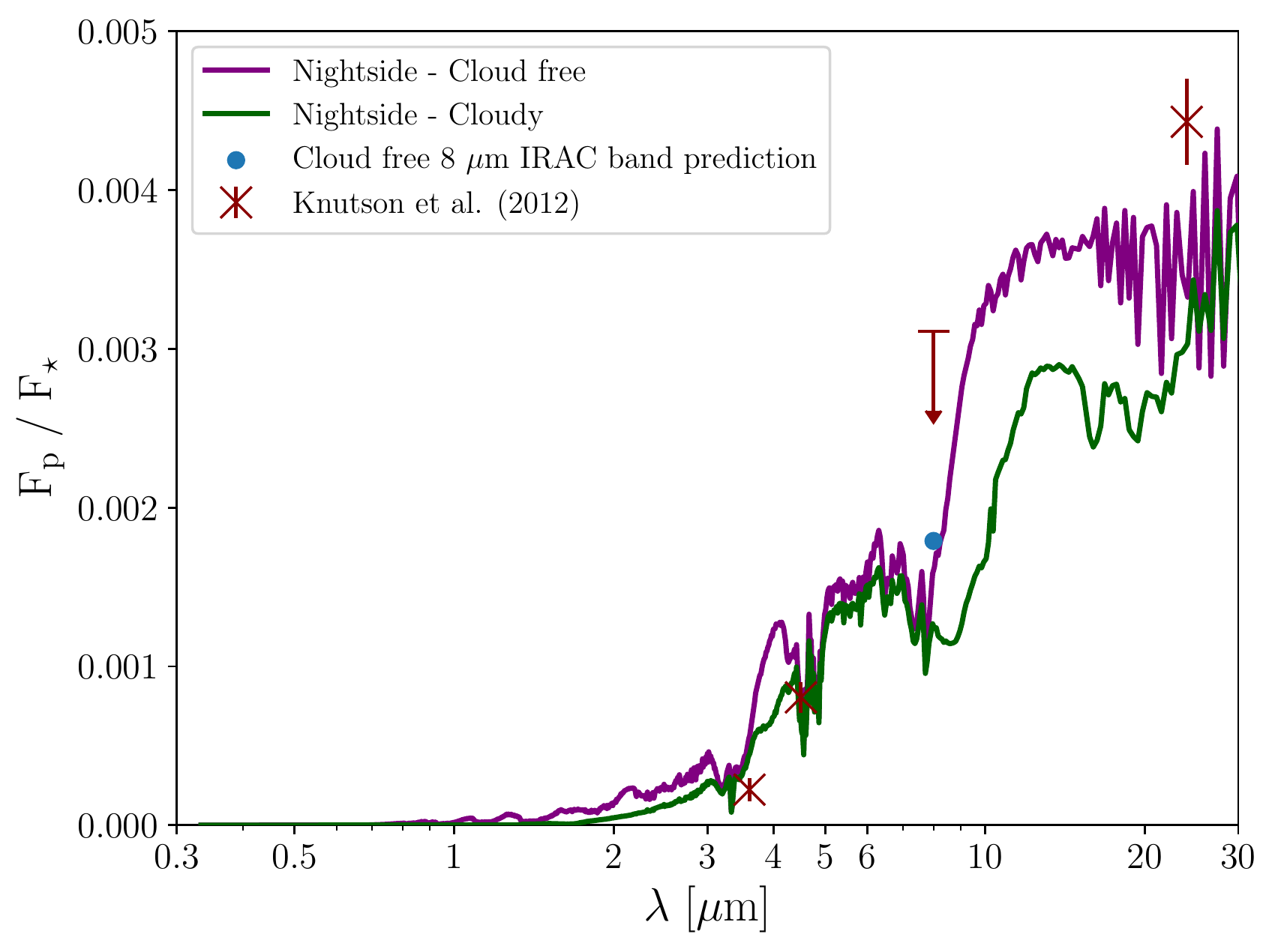}
   \caption{Post-processing of the GCM results from \citet{Lee2016}.
   Top left: transmission spectrum with (blue) and without (orange) cloud opacities compared to the data from \citet{Pont2013} and \citet{McCullough2014}.
   Top right: dayside non-cloudy (blue) and cloudy (orange) emission, nightside non-cloudy (purple) and cloudy (green) emission spectral surface flux density.
   Bottom left: dayside planet/star flux ratio with (blue) and without (orange) cloud opacities compared to the observational data from \citet{Crouzet2014, Barstow2014, Knutson2012} and \citet{Todorov2014}.
   Bottom right: nightside planet/star flux ratio with (purple) and without (green) cloud opacities compared to the Spitzer photometric data from \citet[][and references within]{Knutson2012}.}
   \label{fig:GCM_res}
\end{figure*}

As a 3D spherical grid model, CMCRT is well placed to accurately post-process 3D GCM output.
To test this capability, we post-process output of the cloudy 3D GCM of HD 189733b presented in \citet{Lee2016} using our new methodologies.
We examine the effect of the cloud structures by performing MCRT simulations using the same 3D GCM output with and without cloud opacity.
For input to CMCRT, we use the individual cell temperatures, pressures, cloud properties (mean sizes, number density) and the gas phase element abundances involved in the cloud formation process.
This is not a self-consistent comparison between a cloud-free and cloudy GCM simulation however, since the GCM thermal structures were produced with cloud opacity effects included.
To increase the near-IR wavelength resolution, we use the currently available HELIOS-K produced gas phase opacities in this section, Na and K gas phase opacities are therefore neglected for this model.
In order to focus on the thermal emission of the planet only, we do not include the contribution from reflected stellar light.

For gas phase abundances, the simulation is post-processed assuming chemical equilibrium using \textsc{GGChem} \citep{Woitke2018}, with the local depletion of elements from the cloud formation processes taken from the GCM results.
For the cloud opacity, the simulation is post-processed using effective medium theory and Mie theory the same way as in \citet{Lee2017}.
To simplify the calculation, we assume all cloud opacity and scattering properties at the mean particle radius.
We note this approximation has been investigated by \citet{Powell2018} who showed differences in cloud opacity and scattering properties when considering the integrated effect of a full cloud particle size distribution.
Examining differences in the transmission and emission properties with a full particle size distribution is left to future efforts.

Figure \ref{fig:GCM_res} (top left) presents the transmission spectra results of the GCM post-processing with and without cloud opacity.
The model transmission spectra is compared to published HST \citep{Pont2013, McCullough2014} and Spitzer \citep[][and references within]{Knutson2012} data.
In the infrared, a broad absorption feature from 8 - 30 $\mu$m is produced due to the cloud particles dominant silicate composition at the limbs of the simulated atmosphere \citep[e.g.][]{Wakeford2015}.
The results suggest strong CH$_{4}$ absorption features at $\approx$ 3.3 $\mu$m and 7.8 $\mu$m, however, these features may be modified by non-equilibrium chemistry such as quenching \citep[e.g.][]{Steinrueck2018}, potentially altering the CH$_{4}$ abundance at higher altitudes compared to the local chemical equilibrium assumption in this study.
The GCM transmission spectra produces a flatter optical slope and very muted water features in the near-IR compared to the observational data for the cloudy case.
This suggests that the cloud particles in the upper atmosphere of the GCM are too large compared to what fitting and retrieval to the HD 189733b observational data suggest \citep[e.g.][]{Wakeford2015, Barstow2017}.
A similar conclusions was discussed in \citet{Lines2018, Lines2018b} for microphysical cloud GCM modelling of HD 209458b.

Figure \ref{fig:GCM_res} (top right, bottom left) presents the dayside emission spectra results of the GCM post-processing with and without cloud opacity.
The flux ratio results are compared to published HST \citep{Crouzet2014, Barstow2014} and Spitzer \citep[][and references within]{Knutson2012, Todorov2014} data.
For the parent star luminosities we use a \citet{Castelli2004} \textsc{ATLAS9} stellar atmosphere model with parameters; T$_{\rm eff}$ = 5000 K, $\log$ g = 4.5, [M/H] = 0.0).
The results show that cloud coverage significantly dampens the optical and near-IR emission, generally lowering the emitted flux and increasing the strength of the absorption features across infrared wavelengths.
Our results compare well to the HST data and photometric Spitzer data, however there is an offset between the Spitzer IRS analysis from \citet{Todorov2014}.
This suggests that the photospheric temperature as modelled on the dayside of the GCM may be slightly too high, and a cooler dayside temperature preferred, lowering the peak planetary emission in the Spitzer IRS wavelength range.
A strong emission feature is present at $\approx$ 9.7 $\mu$m, attributed to the Si-O stretching mode from the dominant silicate cloud composition present in the GCM model \citep{Lee2016}.
Due to the strength of this feature, it is quite possible it would have been detectable in the Spitzer IRS observations \citep{Grillmair2008, Todorov2014}.
This suggests that the silicate emission feature is either muted or not present in the real object.
However, this feature arises from the presence of silicates clouds at the temperature inversion regions at higher latitudes in the modelled atmosphere \citep{Lee2016}.
The observation of a cloudy emission feature in an object in the future (e.g. JWST) may be evidence of a temperature inversion in the atmosphere and reveal the cloud composition.

Figure \ref{fig:GCM_res} (top right, bottom right) shows the nightside emission spectra.
The nightside spectra show markedly different features compared to the dayside.
The spectra is largely characterised by the strong CH$_{4}$ absorption at $\approx$ 3.3 $\mu$m and 7.8 $\mu$m.
In contrast to the dayside spectrum, the cloudy nightside produces an absorption feature at $\approx$ 9.7 $\mu$m and 15-20 $\mu$m corresponding to the Si-O stretching and bending modes of the silicates.
This is reminiscent of the absorption feature observed by Spitzer IRS in some L dwarf spectra \citep{Cushing2006}.
The nightside is generally consistent with the 3.6 $\mu$m and 4.5 $\mu$m flux ratio presented in \citet{Knutson2012}, but is not able to reproduce the 8.0 $\mu$m and 24 $\mu$m points.
This suggests the upper atmospheric temperatures in GCM on the nightside may be an underestimate.

\section{Discussion}
\label{sec:discussion}

\subsection{Line wing cutoff}

\begin{figure} 
   \centering
   \includegraphics[width=0.49\textwidth]{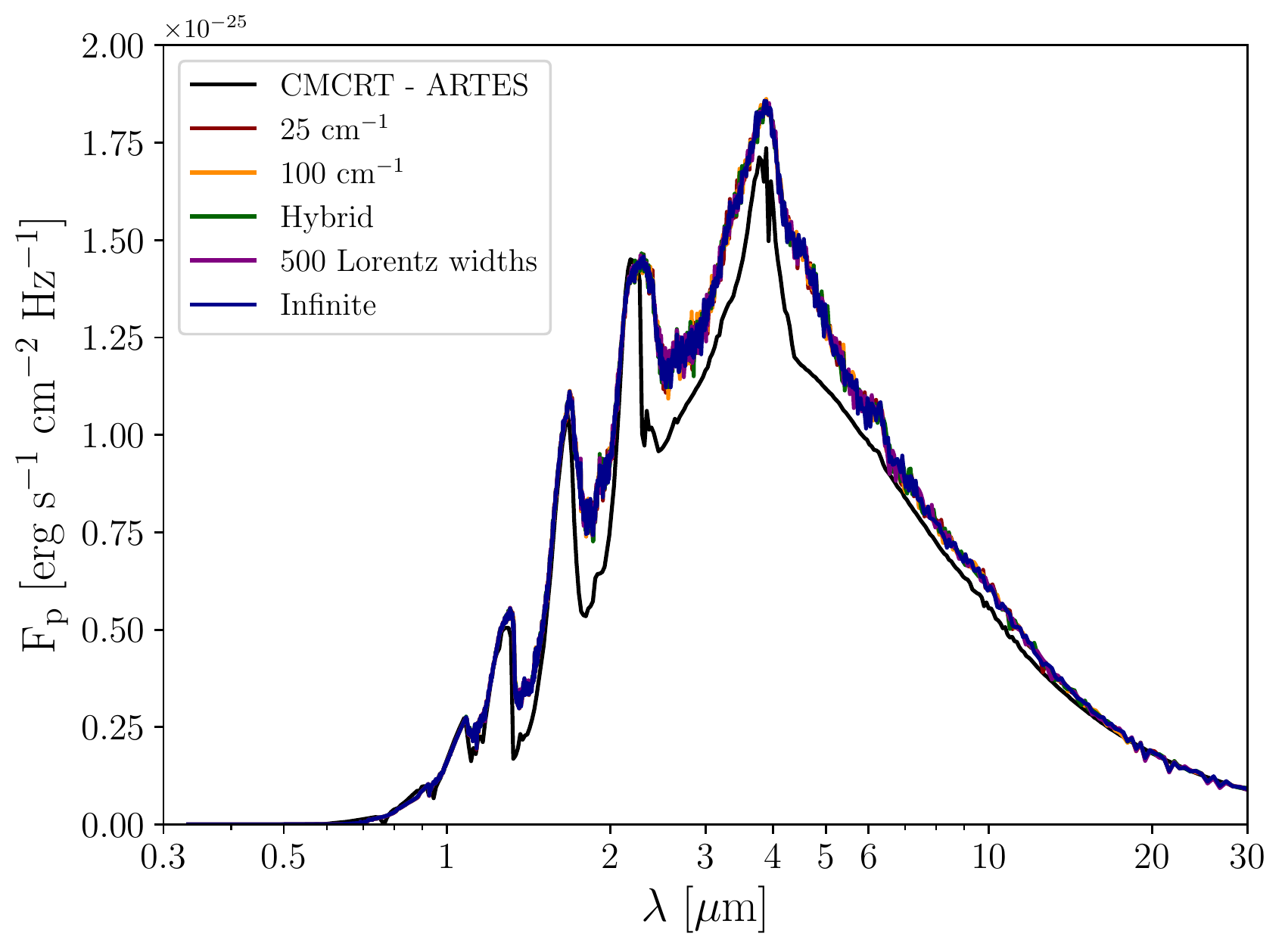}
   \caption{Emission spectra of the T$_{\rm eff}$ = 1500 K benchmark test processed with different line wing treatments.
   Coloured lines shows the results of different treatments for the line wing cutoff when producing the H$_{2}$O opacity tables (see text).
   The black line shows the results using the ARTES opacity tables.}
   \label{fig:1500K_cutoff}
\end{figure}

A possible explanation for the offsets between the CMCRT and the benchmark emission spectra results in Sect. \ref{sec:Stolker} and \ref{sec:Baudino} is the treatment of the input opacity calculation.
The NEMESIS k-tables used in this study applied a Voigt line wing cutoff of 25 cm$^{-1}$, the HELIOS-K tables a cutoff of 100 cm$^{-1}$ and the \citet{Baudino2017} benchmark models use an adaptive cutoff (ATMO) or a sub-Lorentzian lineshape model (petitCODE and Exo-REM).
The ARTES opacities used in \citet{Stolker2017} applied an infinite cutoff (\citet{Min2017}, M. Min priv. corr.).
As shown in \citet{Baudino2017} the choice of line wing cutoff when calculating the input opacities can have significant effects on the end emission spectra results in the forward model.

In order to test the sensitivity of the model output to the line wing cutoff, we produce five H$_{2}$O k-tables with the HELIOS-k code, using the \citet{Barber2006} line list with a variety of line wing cutoff prescriptions found in the literature:
\begin{enumerate}
\item 25, 100 and infinite cm$^{-1}$ absolute cutoff.
\item Hybrid min[25 P (atm), 100] cm$^{-1}$ \citep{Sharp2007}.
\item 500 Lorentz widths \citep{Grimm2015}.
\end{enumerate}
We use the same T-p profile and input quantities as the T$_{\rm eff}$ = 1500 K \citet{Baudino2017} benchmark case, but with only H$_{2}$O as the gas phase absorption opacity.
We also process this T-p profile using the \citet{Stolker2017} opacity table.

Figure \ref{fig:1500K_res} presents the results of this test.
The choice of line wing cutoff between 25 cm$^{-1}$, 100 cm$^{-1}$, Hybrid, 500 Lorentz and infinite widths has a negligible effect on the end emission spectra and produce similar results, which suggests that the H$_{2}$O absorption features are formed at lower pressure p$_{gas}$ $\lesssim$ 1 bar atmospheric layers for this profile, where the pressure broadening effect is less sensitive to the cutoff length.
We can therefore safely conclude that the input opacity treatment is not responsible for the small offsets seen in Sect. \ref{sec:Baudino}.
However, an offset is seen for the results using the \citet{Stolker2017} opacities.
Deeper absorption features are present, as well as a lower peak emission, similar to the benchmark comparison in Sect. \ref{sec:Stolker}.
Possibly, the differences between the models for the \citet{Stolker2017} benchmarks stem from the combination of assuming an infinite cutoff for all gas species rather than just H$_{2}$O considered here, such as that examined in \citet{Baudino2017}.

The issue of the line wing cutoffs and pressure broadening for exoplanet and Brown Dwarf atmospheric opacities are well discussed in the literature \citep[e.g.][]{Sharp2007, Amundsen2014, Grimm2015, Baudino2017} and extensively interrogated in \citet{Hedges2016}.
Future detailed examination of the effects from the choice of line wing cutoff on forward modelling is warranted.

\subsection{3D Geometric Effects}

Since the aim of this paper is to compare a 3D radiative transfer model to contemporary 1D models, it is not surprising that differences would occur in the calculated spectra.
Below we examine our methodologies, and suggest some differences that can lead to variations between our 3D approach and the 1D models.

The comparison to the benchmark transmission spectra produced by NEMESIS (Sect. \ref{sec:NEMESIS}) and \citet{Baudino2017} (Sect. \ref{sec:Baudino}) showed a systematic $\approx$5-15 ppm positive offset between the model output.
In this work, a random starting impact parameter for the ray tracing is chosen on the transit annulus of the planet, in contrast to traditional 1D transmission codes where the ray tracing path is typically taken at the center of the radial cells.
Assuming that an impact parameter width is evenly sampled, simply from spherical geometry, packets randomly starting nearer to the base of the impact parameter width pass through more atmosphere, and hence contribute a slightly higher optical depth in the transmission calculation.
Since the transmission function is given by $T = \exp(-\tau)$, this biases the average transmission for a specific impact parameter width towards higher optical depths, resulting in a larger transit radius compared to the 1D approach.

In some of the emission spectra benchmarks in Sect. \ref{sec:Baudino} a $\approx$5\% additional flux output is seen in CMCRT compared to the 1D models.
This offset is also wavelength dependent, occurring near the peak of the emission flux and the Rayleigh-Jeans tail.
As a model that uses a 3D spherical grid, a fundamental difference to the 1D plane-parallel code is the geometry of the ray tracing through the atmosphere.
Two main approximations break down for plane-parallel methods occurring near the limb of the planet:
\begin{enumerate}
\item The infinite-plane approximation becomes increasingly invalid as the emission angle approaches 90$^{\circ}$, as the sphericity of the planet becomes more important.
\item Near the limbs of the planet, radiation does not emerge from a vertical column of a $\tau$ $\sim$ 1 surface, but a combination of horizontal, low optical depth locations.
\end{enumerate}
As a result of these approximations, the calculated slant path length of a ray to the top of the atmosphere becomes increasingly divergent as the emission angle approaches 90$^{\circ}$.

This effect gives rise to the well known differences in limb darkening behaviour for plane-parallel and spherical models in stellar atmospheres \citep[e.g.][]{Neilson2012}.
Since the T-p profiles used in Sect. \ref{sec:Baudino} are highly isothermal in the upper atmosphere (Fig. \ref{fig:TP}), no limb-darkening effects are seen in the CMCRT output images or expected from the plane-parallel model either.
This suggests that the likely origin of the offsets is directly the difference in path length between the spherical and plane-parallel models, with the 3D model allowing slightly more blackbody emission from the upper atmosphere isothermal regions to escape compared to the 1D approximations.
However, since the offset is prominent in the T$_{\rm eff}$ = 500 K and 1000 K benchmark tests and negligible in the higher temperature tests, it suggests that the magnitude of this effect is dependent on the modelled T-p profile.

Modelling emission near the limbs of tidally locked giant exoplanets may be an important consideration as it has been shown atmospheric jet structures can shift energy away from the sub-stellar point and towards the eastward limb \citep[e.g.][]{Heng2015}.
Extensive cloud structures confined to the westward limb are also theoretically expected to form \citep[e.g.][]{Parmentier2016, Roman2019}, suggesting photon scattering effects near the limb of these exoplanets may also be important to factor into the radiative-transfer problem.
This contrast between the hotter east, and colder and cloudy west terminator regions is well captured by the 3D MCRT model, and will be important for accurate determination of 3D feedback effects of radiative heating and cooling rates inside the atmosphere.

Overall, a rigorous and detailed quantitive comparison between 3D and 1D geometries and the manifestation of these effects in the observable spectra is beyond the scope of this study.
However, such a study is warranted to ensure that radiative-transfer modelling using a variety of different methodologies are well calibrated.

\section{Summary and Conclusions}
\label{sec:conclusion}

We have expanded our 3D Monte Carlo radiative transfer code (Cloudy Monte Carlo Radiative Transfer: CMCRT), originally presented in \citet{Lee2017}, to make use of k-distribution tables and the correlated-k approximation by statistically sampling the g-ordinance weights.
We present a hybrid Monte Carlo and ray tracing method for the calculation of transmission spectra which include a multiple scattering component such as clouds and Rayleigh scattering.
Our method reduces to an equivalent extinction, randomised transit chord algorithm in the absence of a scattering component, allowing the absorption and scattering contributions to the transmission spectra to be individually examined.
We present a MCRT sampling method with k-distributions in emission spectra calculations.
Our application highlights the synergy between the stochastic MCRT model, k-distribution properties and ray tracing methods.

We performed several benchmarking tests in CMCRT for transmission and emission spectra taken from the literature.
Firstly, a direct comparison using identical inputs with the NEMESIS radiative-transfer suite \citep{Irwin2008} compared highly favourably, with differences within 30 ppm for the transmission spectra and 10\% for the emission spectra.
Secondly, we benchmarked CMCRT to the emission spectra results in \citet{Stolker2017}, producing consistent results when using the \citet{Stolker2017} opacity table directly, and offsets between the results when using the k-distribution opacities.
Thirdly, we benchmarked CMCRT to the transmission and emission spectra for prescribed \citet{Guillot2010} T-p profiles to output from the \textsc{ATMO} \citep{Tremblin2015}, \textsc{Exo-REM} \citep{Baudino2015} and \textsc{petitCODE} \citep{Molliere2015,Molliere2017} as presented in \citet{Baudino2017}.
Our chemical equilibrium abundances compared well, with differences arising from thermal dissociation of molecules for the $T_{\rm eff}$ = 2000 K and 2500 K T-p profiles.
Our transmission spectra results generally compare well to the 1D code to within 10s of ppm.
The emission spectra results generally agree, except with a 5-10\% continuum offset at infrared wavelengths in the CMCRT results for the $T_{\rm eff}$ = 500 K and 1000 K T-p profiles.

Lastly, we applied our new methodologies to post-process transmission spectra and emission spectra of the cloudy HD 189733b 3D GCM simulation from \citet{Lee2016}.
Our transmission spectra results with cloud produced too flat an optical slope and near-IR water features, suggesting that the modelled cloud particles in the GCM are too large and at too high altitude.
Our emission spectra results are consistent with the near-IR HST \citep{Crouzet2014, Barstow2014} and Spitzer \citep[][and references within]{Knutson2012} photometric data.
Comparing to the \citet{Todorov2014} Spitzer IRS data suggests that the peak emission on the dayside of the GCM may be too high, indicating a cooler dayside photospheric temperature for the real object.
The presence of an emission feature at the cloud particles stretching and bending modes may be suggestive of an atmospheric temperature inversion where clouds are present, and also reveal the composition of the particles.

We examined the effect of the far line wing cutoff parameter on the emission spectra results, finding negligible differences in our H$_{2}$O tests.
A possible candidate for the differences found between the \citet{Stolker2017} and the CMCRT results is the combination of gas species assuming an infinite cutoff as discussed in \citet{Baudino2017}.

We suggest a 3D spherical geometry effect from randomly sampling impact parameters in CMCRT, which leads to biasing towards a lower transmission through each impact parameter layer, resulting in a slightly higher (ppm level) planetary radius compared to the 1D benchmarks.
We suggest it is likely due to the differences in path lengths between 3D spherical and 1D plane-parallel models at the limb of the simulated planet, a small additional blackbody flux component from the upper atmospheric regions is present in the 3D model compared to the 1D models, the magnitude of which is dependent on the simulated T-p profile.

Despite differences between individual comparisons of output, our benchmarking efforts as a whole suggest our methods produce highly consistent transmission and emission spectra results of comparable quality to the 1D model outputs.
We also conclude from the GCM post-processing results that CMCRT is extremely well suited for accurate simulation of the radiative environment of the complex, 3D inhomogeneous nature of exoplanet atmospheres.
The importance of considering a 3D atmospheric structure on the interpretation of observational and model results is beginning to be explored in detail \citep[e.g.][]{Feng2016, Blecic2017}.
As observational data on these objects increases in quality and quantity, 1D radiative-transfer techniques may quickly become intractable to modelling the 3D spatial complexity of the atmosphere and 3D radiative-transfer techniques more appropriate for the interpretation of the observational data.

This study provides a cautionary tale when comparing between forward model outputs, it can be challenging to diagnose the origin of any systematic difference between models, be it 3D geometrical effects or the individual treatment and parameters used by different groups when calculating input opacities.
Our results also hopefully guide the future of cloud modelling in GCMs, consistently comparing cloudy GCM output to observational data is invaluable for further refinement of GCM modelling techniques and development and testing of cloud formation theories.

\section*{Acknowledgements}
G.K.H. Lee acknowledges support from the University of Oxford and CSH Bern through the Bernoulli fellowship and support from the European community through the ERC advanced grant EXOCONDENSE (PI: R.T. Pierrehumbert).
The participants of the OWL 2018 summer program are thanked for numerous discussions and encouragement on the development of the model.
J. Taylor is a Penrose Scholar and would like to thank the Oxford Physics Endowment for Graduates (OXPEG) for funding this research.
J-L Baudino acknowledges the support of the UK Science and Technology Facilities Council.
T. Stolker and M. Min are thanked for providing data on the ARTES opacities and model output.
V. Parmentier and R.T. Pierrehumbert are thanked for discussions and suggestions on the manuscript content.
Most plots were produced using the community open-source Python packages Matplotlib \citep{Hunter2007}, SciPy \citep{Jones2001}, and AstroPy \citep{Astropy2018}.
The authors' local HPC support at Oxford and CSH Bern is highly acknowledged.




\bibliographystyle{mnras}
\bibliography{bib2} 







\bsp	
\label{lastpage}
\end{document}